\documentclass[twoside,12pt]{article}

\usepackage{epsfig}

\usepackage{graphicx}
\usepackage{amsmath}
\usepackage{amssymb}
\usepackage{array}
\usepackage{wrapfig}
\usepackage{units}
\usepackage{caption}
\usepackage{slashed}
\usepackage{color}

\def\Journal#1#2#3#4{{#1} {#2} (#4) #3 }

\def\NPA{{\em Nucl. Phys.} A}

\def\PLB{{\em Phys. Lett.} B}

\def\PRL{\em Phys. Rev. Lett.}

\def\PREP{\em Phys. Rep.}

\def\PRD{{\em Phys. Rev.} D}
\def\PRC{{\em Phys. Rev.} C}

\newcommand{\dslash}[1]{#1\!\!\!/}

\def\mz{{M_{Z}}}

\newcommand{\be}{\begin{equation}}
\newcommand{\ee}{\end{equation}}
\newcommand{\bea}{\begin{eqnarray}}
\newcommand{\eea}{\end{eqnarray}}

\newcommand{\ba}{\begin{eqnarray}}
\newcommand{\ea}{\end{eqnarray}}

\def\sstw{{\sin^2\theta_W}}
\newcommand{\ket}[1]{\left\lvert #1\right\rangle}
\newcommand{\bra}[1]{\left\langle #1\right\rvert}

\newcommand{\diracslash}[1]{\not\!\! #1}

\newcommand{\nc}{\newcommand}

\nc{\newsection}[1]{\section{#1}\setcounter{equation}{0}}
\nc{\newappendix}[1]{\section*{#1}\setcounter{equation}{0}}
\nc{\scm}{\scriptscriptstyle\mathrm}
\nc{\f}{\frac}
\nc{\baa}{\begin{array}}      \nc{\eaa}{\end{array}}
\nc{\bit}{\begin{itemize}}    \nc{\eit}{\end{itemize}}
\nc{\ben}{\begin{enumerate}}  \nc{\een}{\end{enumerate}}
\nc{\bce}{\begin{center}}     \nc{\ece}{\end{center}}
\nc{\bfl}{\begin{flushright}} \nc{\efl}{\end{flushright}}
\nc{\btb}{\begin{tabular}}    \nc{\etb}{\end{tabular}}
\nc{\eps}{\varepsilon}
\nc{\vp}{\varphi}
\nc{\tvp}{\widetilde{\varphi}}
\nc{\D}{\mbox{$\not\!\!D$}}
\nc{\Db}{\mbox{${\raisebox{2mm}{\boldmath ${}^\leftarrow$}\hspace{-4mm} D}$}}
\nc{\Dfb}{\mbox{$\raisebox{2mm}{\boldmath ${}^\leftrightarrow$}\hspace{-4mm} D$}}
\nc{\vpj }{\mbox{${\vp^\dag i\,\raisebox{2mm}{\boldmath ${}^\leftrightarrow$}\hspace{-4mm} D_\mu\,\vp}$}}
\nc{\vpjt}{\mbox{${\vp^\dag i\,\raisebox{2mm}{\boldmath ${}^\leftrightarrow$}\hspace{-4mm} D_\mu^{\,I}\,\vp}$}}

\def\wt{\widetilde}

\DeclareMathOperator{\Tr}{Tr}

\newcommand\slurp[1]{#1}
\newcommand\sslurp[2]{#1{#2}}
{\catcode`/=\active \expandafter}%
\slurp{\newcommand/}{/\penalty1000\hskip0pt\relax}
{\catcode`:=\active \expandafter}%
\slurp{\newcommand:}{:\penalty1000\hskip0pt\relax}

\newcommand\addspace{\ifcat\nextchar a\spacefactor999. \else.\fi}
{\catcode`\.=\active \expandafter}%
\slurp{\newcommand.}{\futurelet\nextchar\addspace}

\usepackage[unicode]{hyperref} 
\ifx\href\undefined\def\href#1{}\fi
\ifx\texorpdfstring\undefined\def\texorpdfstring#1#2{#1}\fi

\newcommand\myslash{/} \newcommand\mycolon{:}
\newcommand\doi{{\catcode`/=\active \catcode`:=\active \expandafter}\sslurp\realdoi}
{\catcode`/=\active \catcode`:=\active \expandafter}%
\slurp{\newcommand\realdoi[1]{{\let/=\myslash \let:=\mycolon
                               \edef\raw{{http://dx.doi.org/#1}}\expandafter}%
                               \expandafter\href\raw{doi:#1}}}
\newcommand\eprint[2]{{\escapechar-1%
                       \edef\a{\expandafter\string\csname arXiv\endcsname}%
                       \edef\b{\expandafter\string\csname #1\endcsname}%
                       \edef\c{\expandafter\string\csname #2\endcsname}%
                       \edef\d{\noexpand\href{http://arXiv.org/abs/\c}}%
                       \ifx\a\b\expandafter\d\fi{\tt #1:#2}}}
\newcommand\Lbsm{\Lambda}
\newcommand{\lsim}{\buildrel < \over {_\sim}}
\newcommand{\gsim}{\buildrel > \over {_\sim}}

\begin{document}

\title{ \vspace{1cm} 
Low Energy Probes of Physics Beyond the Standard Model
}
\author{
Vincenzo Cirigliano$^{1}$
and 
Michael J. Ramsey-Musolf$^{2,3}$ 
\\
\\
$^1$ Theoretical Division, Los Alamos National Laboratory, \\ Los Alamos, NM 87545, USA
\\
$^2$ 
Department of Physics, University of Wisconsin-Madison,\\
 1150 University Ave.,  Madison, WI 53706, USA
 \\
$^3$ 
California Institute of Technology,\\
 Pasadena, CA 91125, USA}
\maketitle
\begin{abstract} 
Low-energy tests of fundamental symmetries and studies of neutrino properties provide a powerful window on physics beyond the Standard Model (BSM). In this article, we provide a basic theoretical framework for a subsequent set of articles that review the progress and opportunities in various aspects of the low-energy program. We illustrate the physics reach of different low-energy probes in terms of an effective BSM mass scale and illustrate how this reach matches and, in some cases, even exceeds that accessible at the high energy frontier. 
\end{abstract}
%\eject
%\tableofcontents

\newpage

\section{Introduction} 
\label{sec:intro}
The search for physics Beyond the Standard Model (BSM) is now at the forefront of particle physics. The Standard Model (SM) itself represents a triumph of 20th century physics, providing a unified description of three of the known forces of nature, a framework for explaining much of what is observed experimentally, and -- since its inception -- predictions for the existence of new particles such as the top quark that have subsequently been verified. Over the course if its existence, the SM has withstood numerous tests at variety of energy scales, ranging from those associated with atomic processes to energies at the $Z$-pole and above. Indeed, the level of agreement between SM predictions and electroweak precision observables (EWPOs) --  generally at the $10^{-3}$ level or better -- places severe constraints on the BSM mass scale, $\Lbsm$, though the existence of a few significant disagreements may signal the existence of \lq\lq new physics" below this scale.

Despite the impressive successes of the SM, there exist strong observational and theoretical reasons to believe that it is part of a larger framework and that the ultimate goal of BSM searches is to develop a \lq\lq new Standard Model" of nature's fundamental interactions. From an observational standpoint, the experimentally determined components of the \lq\lq cosmic energy budget" given some of the strongest indications of BSM physics. Neither the relic abundance of cold dark matter nor the visible matter-antimatter asymmetry can be explained by the SM. The largest fraction of the energy density of the universe -- the dark energy -- is equally mysterious if not more so\footnote{However, one may account for the acceleration expansion of spacetime with a cosmological constant.}. The observation of neutrino oscillations and that imply non-vanishing masses for the neutrinos require at least a minimal extension of the original version of the SM. And the discrepancy between the muon anomalous magnetic moment, $a_\mu$, as measured by the E821 Collaboration at Brookhaven National Laboratory  and SM expectations -- now at the $\gsim 3\sigma$ level -- could be due to new particles with masses $\lsim\Lbsm$ or below. 

Theoretically, it has long been postulated that the known forces of nature (and possibly others) existed as a single unified force at the moment of the Big Bang, thereby incorporating gravity along with the electroweak and strong interactions in a manner outside the SM. In fact, the running couplings of the SM are suggestive of such a situation, leading to a \lq\lq near miss" for unification at the GUT scale $10^{16}$ GeV (see below) and hinting at the possibility that the presence of additional particles and/or forces would close the gap. Theorists have also contended for some time with the so-called \lq\lq hierarchy problem" associated with quadratically divergent quantum corrections to the Higgs boson mass. The most \lq\lq natural solutions" to this problem point to new physics at the Terascale, though it is not ruled out that $\Lbsm$ is somewhat larger and that some degree of fine-tuning is required. A number of other features of the SM require BSM physics to explain them: the origin of parity-violation in weak interactions; the quantization of electric charge; and values of the input parameters, such as the fermion masses whose spectrum spans eleven orders of magnitude. 

The quest for a new Standard Model that addresses these experimental and theoretical puzzles entails effort at three frontiers: the high energy frontier, now comprised by the CERN Large Hadron Collider; the Cosmological frontier, including probes of the cosmic microwave background and large scale structure as well as indirect astrophysical detection of dark matter; and the Intensity or Precision frontier that is the focus of this issue. The Intensity Frontier (IF) is a itself an interdisciplinary field of research, involving experimental and theoretical physicists from the atomic, nuclear, and high energy physics communities. In the articles that follow, we emphasize the studies within the nuclear physics community, recognizing that there the lines between these communities are not precisely defined. That being said, we distinguish three classes of studies being pursued by members of the nuclear physics community:
\begin{itemize}
\item[(a)] {\em Rare and forbidden processes.} Highly suppressed or strictly forbidden in the Standard Model, these observables include permanent electric dipole moments (EDMs) of leptons, nucleons, and atoms; the neutrinoless double $\beta$-decay of nuclei ($0\nu\beta\beta$); and processes that do not conserve flavor, 
such as the charged lepton flavor violating (CLFV) decay $\mu\to e\gamma$. In most cases, the observation of such an observable would provide \lq\lq smoking gun" evidence for BSM physics, though in the case of EDMs of systems containing quarks the so-called \lq\lq $\theta$-term" in the QCD Lagrangian could be responsible.
\item[(b)] {\em Precision tests.} These studies measure quantities that are allowed within the Standard Model, but do so at a level of precision that could uncover tiny deviations from Standard Model expectations and signal the presence of BSM dynamics. Of particular interest to the nuclear physics community are the weak decays of the muon and hadrons containing light quarks (including heavy nuclei), parity-violating asymmetries in the scattering of polarized electrons from unpolarized targets, and the (anomalous) magnetic moments of the muon and of neutrinos.  The interpretation of these measurements in terms of BSM physics depends critically on the reliability of Standard Model predictions as well as the level of experimental precision.
\item[(c)] {\em Cosmological and astrophysical probes.} The production of neutrinos in astrophysical processes, such as energy production in the sun or supernovae, can provide unique clues about the underlying fundamental physics -- as can the analysis of the role played by neutrinos in cosmological processes, such as those responsible for large scale structure. At the other end of the distance scale spectrum, terrestrial experiments exploiting heavy nuclei as detectors give the most sensitive means for detecting the presence of relic dark matter in many scenarios. In both cases, obtaining a sufficiently reliable understanding of the \lq\lq laboratories" ({\em e.g.}, the Sun or heavy nuclei) is important for a proper interpretation of these studies in terms of possible BSM dynamics. 
\end{itemize}

In the remainder of this issue, we discuss in detail the theoretical framework for the interpretation of these studies. Our goal is not to review the state of the field experimentally, but rather to provide a theoretical context in which to view the significance of various measurements. This endeavor is particularly timely, given the influx of new results from other frontiers that clearly impacts the significance and interpretation of the IF studies. Indeed, given the dynamic nature of BSM searches, we will not attempt to provide definitive interpretations regarding particular BSM scenarios, but rather to provide a context for the on-going evaluation of the implications of nuclear physics fundamental symmetry tests and  neutrino property studies. In this respect, we view this Issue as a providing the background for a working group website  that will contain up-do-date information about key experimental results and their interpretation. 

The articles that follow are organized according to the outcome of the workshop \lq\lq Beyond the Standard Model in Nuclear Physics", held at the University of Wisconsin-Madison in October, 2011 and supported in part by the U.S. Department of Energy Office of Science Nuclear Physics Program and the corresponding office at the National Science Foundation. The workshop was attended by roughly twenty senior and junior theorists from North America, with discussions breaking down into various topical working groups along the lines of the follow articles. 

In this introductory article, we first review the Standard Model and its renormalization as applied to the processes of interest here. We subsequently cast the possible effects of BSM physics in the context of effective operators containing the Standard Model fields and possibly a (light) right-handed neutrino. For most purposes, it suffices to concentrate on operators of mass dimension six or lower, whose Wilson coefficients encode the effects of new dynamics at the BSM scale $\Lambda$ taken to be well above the weak scale $v=246$ GeV. 
We conclude with an overview of the discovery potential and diagnosing power of the various probes discussed in depth in the remainder of this volume, 
emphasizing  in each case the theoretical  challenges.  

%We subsequently provide a few model illustrations, incorporating the constraints from the LHC as of this writing. We conclude with an introductory discussion of an important, recurring issue for the interpretation of low-energy precision symmetry tests: the interplay of strong and electroweak interactions. On the one hand, the effects of non-perturbative strong interactions must be sufficiently tamed theoretically in order to interpret a given measurement in terms of BSM physics. Perhaps, the most well-known example entails hadronic vacuum polarization and light-by-light scattering contributions to $a_\mu$. On the other hand, electroweak weak processes may be used to probe novel aspects of the non-perturbative strong interaction, as in the recently completed program of parity-violating electron scattering measurements that have yielded stringent upper bounds on strange quark contributions to the nucleon's electromagnetic structure. In this issue, we highlight one on-going field of study along these lines, namely, parity-violation in purely hadronic interactions. 

Before proceeding, we refer the reader to other recent reviews of these topics, including an earlier article in this journal \cite{Erler:2004cx}; a more recent application to supersymmetry \cite{RamseyMusolf:2006vr}; and reviews of EDMs\cite{Pospelov:2005pr}, $0\nu\beta\beta$\cite{Elliott:2002xe,Elliott:2004hr,Avignone:2007fu}, ordinary $\beta$-decay \cite{Herczeg:2001vk,Severijns:2006dr,Severijns:2011zz}, weak neutral currents\cite{Kumar:2013yoa,Musolf:1993tb} as well as hadronic parity-violation \cite{Adelberger:1985ik,RamseyMusolf:2006dz}. Some of what follows constitutes an update of those earlier works, though the articles in this Issue represent a broader and more thorough treatment.

\section{The Standard Model and its Renormalization}
\subsection{\it The Lagrangian \label{sec:SM}}

As a renormalizable theory, the SU(3)$_C\times$SU(2)$_L\times$U(1)$_Y$  Standard Model consists entirely of operators containing mass dimension $d=4$ that appear in the Lagrangian
\be
\mathcal{L}_\mathrm{SM} = \mathcal{L}_\mathrm{gauge}+\mathcal{L}_\mathrm{matter}+\mathcal{L}_\mathrm{Higgs}
\ee
where 
\be
\mathcal{L}_\mathrm{gauge}=-\frac{1}{2}\Tr\left(G_{\mu\nu} G^{\mu\nu}\right) -\frac{1}{2}\Tr\left(W_{\mu\nu} W^{\mu\nu}\right)-\frac{1}{4} B_{\mu\nu} B^{\mu\nu}\ \ \ ,
\ee
with $G_{\mu\nu}=T^A_3 G_{\mu\nu}^A$,  $W_{\mu\nu}=T^a_2 W_{\mu\nu}^a$, and $B_{\mu\nu}$ being the SU(3)$_C$, SU(2)$_L$, and U(1)$_Y$ field strength tensors (expressed in terms of the generators $T^a_j$), and 
\bea
\label{eq:matter}
\mathcal{L}_\mathrm{matter}&=&\sum_{f=q,u,d,\ell,e}\ {\bar f} {\not\!\! D} f +\mathcal{L}_Y\\
\mathcal{L}_\mathrm{Higgs} & = & \left(D_\mu \varphi \right)^\dag D^\mu \varphi - V(\varphi)\ \ \ ,
\eea
where the sums run over the left handed quark (lepton) S(2)$_L$ doublets $q$ ($\ell$) and right handed quark (lepton) singlets $u,d$ ($e$)
\be
l^i =
\left(
\begin{array}{c}
\nu_L^i\\
e_L^i
\end{array}
\right)
\qquad e^i = e_R^i
\qquad q^i =
\left(
\begin{array}{c}
u_L^i \\
d_L^i
\end{array}
\right)  \qquad
u^i = u_R^i \qquad
d^i = d_R^i~,
\label{eq:fermions}
\ee
and the Higgs doublet $\varphi$
\be
\varphi =
\left(
\begin{array}{c}
\varphi^+ \\
\varphi^0
\end{array}
\right)~,
\ee
having a potential energy given by $V(\varphi)$. The covariant derivatives are given by 
\be
D_\mu = \partial_\mu-ig_3\sum_{A=1}^8 T^A_3 G_\mu^A-ig_2 \sum_{a=1}^3T^a_3 W_\mu^a - ig_1 T_1 B_\mu
\ee
with generators $T^A_1=\lambda^A/2$ (acting on quarks only),  $T^a_2=\sigma^a/2$, and  $T_1=Y/2$ expressed in terms of the Gell-Mann matrices $\lambda^A$, Pauli matrices $\sigma^a$, and hypercharge $Y$ (the gauge couplings are alternately denoted $g_s\equiv g_3$, $g\equiv g_2$, and $g^\prime\equiv g_1$). The Yukawa Lagrangian responsible for mass generation through electroweak symmetry-breaking (EWSB) is
\be
\mathcal{L}_Y = -Y_u^{jk} {\bar q}_j \epsilon \varphi^\ast u_k-Y_d^{jk} {\bar q}_j  \varphi d_k-Y_\ell^{jk} {\bar \ell}_j \varphi e_k\ \ \ ,
\ee
where repeated indices denote a sum over fermion families. Note that we have not included terms responsible for neutrino mass (see below). After EWSB, $\bra{0} \varphi^T=\ket{0} = (0, v/\sqrt{2})$ and diagonalization of the Yukawa matrices $Y_f^{jk}$ through the unitarity transformations
\be
\label{eq:flavrot}
u_L^j = \left[ S_u\right]_{jk} U_L^k\ , \qquad d_L^j = \left[ S_d\right]_{jk} D_L^k\ , \qquad u_R^j = \left[ T_u\right]_{jk} U_R^k\ , \qquad d_R^j = \left[ T_d\right]_{jk} D_L^k\ \ \ \mathrm{ etc.}
\ee
the flavor diagonal fermion mass matrices  are given by
\be
m_f=\frac{v}{\sqrt{2}}\ S_f^\dag Y_f T_f\ \ \ .
\ee
After the flavor rotations (\ref{eq:flavrot}) the neutral current gauge interactions in Eq.~(\ref{eq:matter}) remain flavor diagonal, while the charge changing interactions have the form
\be
\mathcal{L}_{CC} = -\frac{ig_2}{\sqrt{2}} V_\mathrm{CKM}^{jk} {\bar U}_L^j \diracslash{W}^+ D_L^k +\mathrm{h.c.}\ \ \ ,
\ee
with $W_\mu^\pm$ being the charged weak gauge boson fields and
\be
V_\mathrm{CKM} = S_u^\dag S_d
\ee
being the Cabibbo-Kobayashi-Maskawa matrix. 

The incorporation of non-vanishing neutrino masses as implied by the observation of neutrino oscillations, can be minimally accomplished through the addition to $\mathcal{L}_\mathrm{SM}$ of either a dimension-four Dirac mass term or a 
higher dimensional Majorana mass term (or both).
The Dirac mass term reads
\be
\label{eq:mnudirac}
\mathcal{L}^\nu_\mathrm{Dirac} = - Y_\nu^{jk} {\bar \ell}_j \epsilon \varphi^\ast \nu_{R\, k} ~, 
\ee
where $\nu_{R\, k}$ is a right-handed neutrino Dirac spinor field and where 
the non-vanishing $m_\nu$ arises when $\varphi$ obtains its vev. 
Like all of the operators appearing in $\mathcal{L}_\mathrm{SM}$, the Dirac mass term in Eq.~(\ref{eq:mnudirac}) has mass dimension four and does not modify the renormalizability of the SM. 
For $v=246$ GeV, the Dirac Yukawa couplings $Y_\nu$ must have a magnitude $\lsim 10^{-12}$ to be consistent with the upper bounds on the scale of neutrino masses obtained from cosmology.
We will discuss the Majorana mass term in the next section, 
since it  is related to a dimension-five non-renormalizable term in 
the effective Lagrangian that parameterizes  physics beyond the Standard Model.

\subsection{\it Renormalization \label{sec:renorm}}

A key consideration in the interpretation of low-energy BSM probes is the impact of electroweak radiative corrections on precision observables. When interpreting measurements of electroweak processes performed with $10^{-3}$ precision or better, reliance on the tree-level SM Lagrangian is not sufficient, as electroweak radiative corrections generically arise at the $\mathcal{O}(\alpha/\pi)$ level. Moreover, in multiple cases the corrections themselves sample momenta at or below the hadronic scale, rendering them susceptible to strong interaction effects and the attendant uncertainties. Consequently, one must devote considerable care in evaluating these corrections and assessing the overall level of theoretical uncertainty. In order to be confident that any deviation from a SM prediction is a {\em bona fide} indicator of BSM physics, one must have in hand a sufficiently reliable SM computation. Among the better-known examples where strong-interaction effects have challenged the theoretical community are the hadronic vacuum polarization (HVP) and hadronic light-by-light (HLBL) contributions to $a_\mu$; the $W\gamma$ box graph to neutron and nuclear $\beta$-decay;  the $Z\gamma$ box graph contribution to the PV asymmetries in elastic $ep$ scattering; and the hadronic contributions to the running of $\sstw$. In contrast, these issues hold less concern for the interpretation of rare or forbidden processes, as the SM contributions are already sufficiently small to be negligible. 

Apart from its importance for obtaining sufficiently reliable SM predictions,  the analysis of radiative corrections can also be vital for the assessment of BSM contributions. In the case of SUSY, for example, the imposition of R-parity conservation implies that superpartners contribute to the observables of interest here only via loops. Moreover, if their masses are sufficiently light, then the effective operator framework described in Section \ref{sec:BSM} may not be appropriate, as one cannot integrate out these new degrees of freedom at a scale well above the weak scale. In addition, precision measurements at the Z-pole and above constrain various combinations of these loop effects, so one must consistently incorporate these constraints when analyzing the prospective impacts on low-energy processes.

With these considerations in mind, we provide here a brief overview of renormalization in the SM, introducing some formalism that will be of use throughout the remainder of this Issue. In doing so, we largely follow the discussion of Ref.~\cite{RamseyMusolf:2006vr}. We generally utilize dimensional regularization with the modified minimal subtraction scheme (${\overline {\rm MS}}$), though the reader should be aware that other schemes (such as on-shell renormalization) are often adopted in the literature\footnote{Note that in the case of SUSY, one must employ a variation of ${\overline {\rm MS}}$ in order to maintain supersymmetry at the loop level. The regulator in this case is dimensional reduction,  (DR), wherein one works in $d=4-2\varepsilon$ spacetime dimensions while retaining the Clifford algebra appropriate to fermion field operators in $d=4$ dimensions. The corresponding renormalization scheme, analogous to ${\overline {\rm MS}}$, is known as modified dimensional reduction, or ${\overline {\rm DR}}$.}. Renormalized quantities are obtained by introducing counterterms  that remove the factors of $1/\varepsilon-\gamma+\ln 4\pi$ that arise in divergent one-loop graphs. All ${\overline {\rm MS}}$-renormalized (finite) quantities will be indicated by a hat, as in $\mathcal{O}\to\widehat{\mathcal{O}}$.

Most low energy precision electroweak observables of interest to nuclear physics are mediated at lowest order by the exchange of a virtual gauge boson (GB), so we consider first the renormalization of GB propagators and GB-fermion interactions. We first discuss renormalization relevant to charged current (CC) processes in order to introduce notation and conventions and subsequently discuss the neutral current (NC) sector.

\subsection{\it Charged Current Processes}

Radiative corrections to CC amplitudes naturally divide into four topologies: (a) $W$-boson propagator corrections; (b) corrections to the $W$-fermion vertices; (c) fermion propagator corrections; and (d) box graphs. There exist extensive studies of these corrections in the SM, dating back to the seminal work of Refs.~\cite{Sirlin:1977sv} and the subsequent analysis of Refs.~\cite{Marciano:1985pd,Marciano:1993sh,Marciano:2005ec,Ivanov:2012qe}. We refer the reader to these studies and references therein for additional details.
% Illustrative contributions in the SM  are indicated in Fig.~\ref{fig:cccorr} [{\bf place holder}]

%\begin{figure}[ht]
%\begin{center}
%\resizebox{6 in}{!}{
%\includegraphics*[20,160][580,620]{cccorrection.ps}}
%\caption{Representative supersymmetric corrections to charged current observables: (a) $W$-boson propagator corrections; (b) vertex corrections; (c) external leg corrections; and (d) box graph contributions.}
%\label{fig:cccorr}
%\end{center}
%\end{figure}

One loop corrections to the $W$-boson propagator, fermion propagator, and $W$-fermion vertices are divergent. After renormalization, the W-boson propagator, $iD_{\mu\nu}(k)$ takes the general form in the Feynman gauge
\be
iD_{\mu\nu}(k) = -i\left[T_{\mu\nu}{\hat D}_{WW}^T(k^2)+L_{\mu\nu}{\hat D}_{WW}^L(k^2)\right]
\ee
where the transverse and longitudinal projection operators are given by
\bea
T_{\mu\nu} & = & -g_{\mu\nu}+ k_\mu k_\nu/k^2 \\
L_{\mu\nu} & = & k_\mu k_\nu/k^2
\eea
and ${\hat D}_{WW}^{T,L}(k^2)$ are finite scalar functions. For the low-energy processes of interest here, effects associated with the longitudinal term are suppressed by light fermion masses, so we will not discuss the component further. The renormalized transverse component is given by
\be
\left[ {\hat D}_{WW}^T(k^2)\right]^{-1} = k^2-{\hat M}_W^2 +{\hat\Pi}_{WW}^T(k^2)\ \ \ .
\ee
Here ${\hat M_W}$ is the finite part of the bare W-boson mass parameter appearing in the renormalized Lagrangian after electroweak symmetry breaking and ${\hat\Pi}_{WW}^T(k^2)$ gives the finite loop contribution. Both ${\hat M_W}$ and ${\hat\Pi}_{WW}^T(k^2)$ depend on the t'Hooft renormalization scale $\mu$. The physical W-boson mass is $\mu$-independent and is defined by the value of $k^2$ giving $[{\hat D}_{WW}^T(k^2
=M_W^2)]^{-1}=0$, {\em i.e.}, the pole of the propagator. 
%The finite residue ${\hat Z}_W$ of the pole is
%\be
%\label{eq:Wres}
%{\hat Z}_W = \left[1+{\hat\Pi}_{WW}^{T\ \prime}(M_W^2)\right]^{-1}\ \ \ .
%\ee

The corresponding expression for the renormalized, inverse fermion propagator is 
\be
{\hat S}_f^{-1}(k) = \dslash{k}-{\hat m}_f +\left[ {\hat A}_L(k^2)\dslash{k} +{\hat B}_L(k^2)\right] P_L 
+ \left[ {\hat A}_R(k^2)\dslash{k} +{\hat B}_R(k^2)\right]P_R
\ee
where $P_{L,R}$ are the left- and right-handed projectors and the ${\hat A}_{L,R}$ and ${\hat B}_{L,R}$ contain the finite loop contributions.  The physical fermion mass is given by 
\be
m_f=\left[ {\hat m}_f -\frac{1}{2}{\hat B}_L(m_f^2)
-\frac{1}{2}{\hat B}_R(m_f^2)\right]\,  \left[1+\frac{1}{2}{\hat A}_L(m_f^2)  + \frac{1}{2}{\hat A}_R(m_f^2) \right]^{-1}\ \ \ ,
\ee
while the residue of the pole is
\be
{\hat Z}_\psi = \left[1+{\hat A}_L(m_f^2) P_L + {\hat A}_R(m_f^2) P_R\right]^{-1} \ \ \ .
\ee
%Note that for CC interactions in the SM,  the left-handed (LH) components given the dominant contribution to physical amplitudes, as the presence of right-handed (RH) components will be suppressed by factors of the fermion masses\footnote{For example, the weak magnetic moment operator in the SM is chirality-odd and  is generated by one-loop vertex corrections that contain single insertions of the Yukawa interaction.}.  

The renormalized vertex functions for CC amplitudes are relatively straightforward. We illustrate using the muon decay process  $\mu^-\to \nu_\mu W^-$, for which the tree-level amplitude is 
\be
i{\cal M}_{0}^{W\mu\nu_\mu} = i \frac{g}{\sqrt{2}}{\bar \nu }_\mu \diracslash{W}^{\ +} P_L \mu
\ee
After one-loop renormalization, one has
\be
i{\cal M}_{0}^{W\mu\nu_\mu }+i{\cal M}_{\rm vertex}^{W\mu\nu_\mu } =i \frac{{\hat g}(\mu)}{\sqrt{2}}\left[1+{\hat F}_V(k^2)-\frac{1}{2}\left\{
{\hat A}_L^\mu(m_\mu^2)+{\hat A}_L^{\nu_\mu}(0)\right\}\right] {\bar \nu }_\mu \diracslash{W}^{\ +} P_L \mu
\ee
where ${\hat g}(\mu)$ is the running SU(2)$_L$ gauge coupling and ${\hat F}_V(k^2)$ is the finite part of the one-loop vertex correction. 
%We have not included the effect of $W$-boson wavefunction renormalization [given from Eq.~(\ref{eq:Wres})]. Because we will  embed this renormalized vertex into the four-fermion amplitude obtained via the exchange of a virtual $W$, use of the renormalized $W$ propagator is appropriate. 

For processes such as  $\mu$-decay and $\beta$-decay, one requires the renormalized four-fermion amplitude, ${\cal M}_{\rm box}^{\rm CC}$. In addition to the vertex and propagator corrections introduced above, ${\cal M}_{\rm box}^{\rm CC}$ receives additional finite one-loop contributions associated with box graphs involving the exchange of two vector bosons. Since the external fermion masses and  momenta for nuclear physics processes are small compared to the weak scale, the box contributions  have the form of a product of two left-handed currents, $(V-A)\otimes(V-A)$.  In the case of $\mu^-\to \nu_\mu e^-{\bar\nu}_e$ one has
\be
i{\cal M}_{\rm box}^{\rm CC} = -i \frac{{\hat g}^2}{2{\hat M}_W^2}{\hat \delta}_{\rm box} \ {\bar \nu}_\mu  \gamma^\lambda P_L \mu \ {\bar {e } }\gamma_\lambda P_L {\nu_{\bar e}} + \cdots\ \ \ \ ,
\ee
where the $+\cdots$ indicate terms whose structure differs from the $(V-A)\otimes(V-A)$  structure of the tree-level CC amplitude. In the SM, such terms will be suppressed by factors of $m_\mu^2/M_W^2$.

Including the box contribution along with the other renormalized one-loop contributions and working in the $k^2 << M_W^2$ limit, one obtains the renormalized four fermion amplitude:
\bea
i{\cal M}_{\rm tree}^{\rm CC}&+&i{\cal M}_{\rm vertex}^{\rm CC}+i{\cal M}_{\rm propagator}^{\rm CC}+i{\cal M}_{\rm box}^{\rm CC} = 
-i\frac{{\hat g}^2}{2{\hat M}_W^2}\Bigl[1+\frac{{\hat\Pi}_{WW}^T(0)}{{\hat M}_W^2}  \\
\nonumber
& -& \frac{1}{2}\left\{
{\hat A}_L^\mu(m_\mu^2)+{\hat A}_L^e(m_e^2)+{\hat A}_L^{\nu_e}(0)+{\hat A}_L^{\nu_\mu}(0)\right\}
+{\hat F}_V^e(0)+{\hat F}_V^\mu(0)+{\hat \delta}_{\rm box}\Bigr] \\
\nonumber
&&\times \ {\bar \nu }_\mu \gamma^\lambda P_L \mu \ {\bar {e } }\gamma_\lambda P_L {\nu_{\bar e}} +\cdots \ \ \ ,
\eea
or
\be
i{\cal M}_{\rm one-loop}^{\rm CC}=-i\frac{{\hat g}^2}{2{\hat M}_W^2}\left[1+\frac{{\hat\Pi}_{WW}^T(0)}{{\hat M}_W^2}+{\hat\delta}_{VB}\right]{\bar \nu }_\mu \gamma^\lambda P_L \mu \ {\bar {e } }\gamma_\lambda P_L {\nu_{\bar e}} +\cdots \ \ \ ,
\ee
where ${\hat\delta}_{VB}$ denotes the fermion propagator, vertex, and box graph contributions. 

The example of the muon decay amplitude is particularly important, since the measurement of the muon lifetime provides one the three required inputs for the SM gauge-Higgs sector. Taking into account the one-loop corrections and including the bremstraahlung contribution leads to the muon decay rate:
\bea
\label{eq:taumu}
\frac{1}{\tau_\mu} & = & \frac{m_\mu^5}{96\pi^3}\left(\frac{{\hat g}^2}{8{\hat M}_W^2}\right)^2  \left[1+\frac{{\hat\Pi}_{WW}^T(0)}{{\hat M}_W^2}+{\hat\delta}_{VB}\right]^2 \  + \ {\rm brem} \\
\nonumber
& = & \frac{m_\mu^5}{192\pi^3} G_\mu^2\left[1+\delta_{\rm QED}\right] \ \ \ ,
\eea
where $\tau_\mu$ is the muon lifetime and the second equality defines the $\mu$-decay Fermi constant, $G_\mu$, and where 
\be
\delta_{\rm QED} = \frac{\alpha}{2\pi} \left( \frac{25}{4}-\pi^2 \right)+\cdots 
\ee
denotes the contributions from real and virtual photons computed in the Fermi theory of the decay.
As a result, one may express the $\mu$-decay Fermi constant in terms of the SU(2$)_L$ coupling, $W$-boson mass parameter, and radiative corrections:
\be
\frac{G_\mu}{\sqrt{2} }= \frac{{\hat g}^2}{8{\hat M}_W^2}\left[1+\frac{{\hat\Pi}_{WW}^T(0)}{{\hat M}_W^2}+{\hat\delta}_{VB}^{(\mu)}\right] \equiv \frac{{\hat g}^2}{8{\hat M}_W^2}\left(1+{\Delta \hat r}_\mu\right) \ \ \ ,
\ee
where ${\hat\delta}_{VB}^{(\mu)}$ is given by ${\hat\delta}_{VB}$ but with the Fermi theory QED contributions subtracted out. 

Along with the fine structure constant and the Z-boson mass, $M_Z$, the value of $G_\mu$  is one of the three most precisely determined parameters in the gauge-Higgs sector of the SM. When computing other electroweak observables, it is conventional to express ${\hat g}^2$ in terms of $G_\mu$, ${\hat M}_W$, and the correction $\Delta \hat r_\mu$:
\be
\label{eq:ghat}
{\hat g}^2 = \frac{8{\hat M}_W^2 G_\mu}{\sqrt{2}}\ \frac{1}{1+{\Delta \hat r}_\mu}\ \ \ .
\ee
To illustrate the use of this expression, consider now the corresponding amplitude for the $\beta$-decay $d\to u e^-{\bar\nu}_e$:
\bea
\label{eq:betaampl1}
i{\cal M}_{\beta-{\rm decay} } &=& i\frac{{\hat g}^2}{2{\hat M}_W^2}\, V_{ud}\, \left(1+{ \Delta \hat r}_\beta\right) \, {\bar u}
\gamma^\lambda P_L \, d\  {\bar e }\gamma_\lambda P_L {\nu_{\bar e}} \ \ \ ,
\eea
where ${ \Delta \hat r}_\beta$ is the analog of ${ \Delta \hat r}_\mu$ for the semileptonic four-fermion amplitude but, in contrast with the $\mu$-decay case, also contains virtual photon contributions. Infrared divergences associated with the latter are cancelled by real radiation contributions to the decay rate. 
Substituting ${\hat g}^2$ as given in Eq.~(\ref{eq:ghat}) we then obtain
\bea
\label{eq:betaampl2}
i{\cal M}_{\beta-{\rm decay} }&=& i \frac{G_\mu}{\sqrt{2}}\, V_{ud}\, \left(1+{ \Delta \hat r}_\beta-{ \Delta \hat r}_\mu\right)
{\bar u} \gamma^\lambda (1-\gamma_5)\, d\  {\bar e} \gamma_\lambda (1-\gamma_5)  {\nu_{\bar{e}}}\ \ \ .
\eea
Importantly, the SM prediction for ${\cal M}_{\beta-{\rm decay} }$ depends on the difference of the purely leptonic corrections ${\Delta \hat r}_\mu$ and the semileptonic corrections ${\Delta \hat r}_\beta$. Any corrections that are common to both processes, such as corrections to the $W$-boson propagator or to the first-generation lepton external legs, will cancel in this difference. 

Looking ahead to the review of weak decays~\cite{charged-current}, 
we alert the reader to some notational differences. In that work one encounters the quantities $\delta_{\beta}$, $\delta_\mu$,
$\epsilon_L$, and $\epsilon_\mu$. The correspondence with the foregoing discussion is: $\Delta {\hat r}_{\beta,\, \mu}\to\delta_{\beta,\, \mu}$ while $\epsilon_L$ and $\epsilon_\mu$ denote the corresponding contributions from BSM physics that enter, respectively, like ${\Delta \hat r}_\beta$ and ${\Delta \hat r}_\mu$.

\subsection{\it Neutral Current Processes}

Although renormalization of neutral current (NC) amplitudes is similar to that of CC interactions, new aspects arise associated with mixing between the SU(2)$_L$ and U(1)$_Y$ sectors. Among the earliest studies of NC renormalization that addressed these issues are those of Refs.~\cite{Marciano:1980pb,Marciano:1982mm,Sarantakos:1982bp,Marciano:1983ss}. Here, we highlight three features: the presence of right-handed as well as left-handed fermion fields; the relative normalization of the NC and CC amplitudes encapsulated by an appropriately defined \lq\lq $\rho$-parameter"; and the appearance of the weak mixing angle whose scale-dependence (in the ${\overline {\rm MS}}$) arises from loop effects. 

To illustrate this added richness, consider the general structure of the renormalized amplitude for the neutral current process $\ell+f\to\ell +f$ is
\begin{eqnarray}
\label{eq:nconeloop}
\lefteqn{i{\cal M}_{\rm one - loop}^{\rm NC} =} \nonumber \\
&& -i\frac{G_\mu}{2\sqrt{2}} {\hat \rho}_{\rm NC}(k^2) \frac{M_Z^2}{k^2-M_Z^2+i M_Z\Gamma_Z} {\bar\ell}\,  \gamma^\lambda({\hat g}_V^\ell+{\hat g}_A^\ell \gamma_5)\, \ell \ {\bar f}\, \gamma_\lambda({\hat g}_V^f+{\hat g}_A^f \gamma_5)\,  f + {\rm box},
\end{eqnarray}
where $\ell$ and $f$ denote the lepton and fermion spinors, respectively, and 
``+ box" denotes the box diagram contributions.  The quantity  $\hat\rho_{\rm NC}$ is a normalization factor common to all four-fermion NC processes that can be expressed in terms of gauge boson masses, the ${\hat \Pi}^T_{VV}(k^2)$, and ${\Delta \hat{r}}_\mu$ 
\cite{veltman}:
\be
\label{eq:rhonc1}
\hat\rho(k^2)_{\rm NC} =1+
\frac{{\rm Re}\ {\hat\Pi}_{ZZ}^T(M_Z^2)}{M_Z^2}-\frac{{\hat\Pi}_{WW}^T(0)}{M_W^2}
-\frac{\textrm{Re}\, \left[{\hat\Pi}_{ZZ}^T(k^2)-{\hat\Pi}_{ZZ}^T(M_Z^2)\right]}{k^2-M_Z^2} -{\hat\delta}_{VB}^{(\mu)}\ \ \ ,
\ee
where
\be
\label{eq:mzhat}
M_Z^2={\hat M}_Z^2-{\hat\Pi}_{ZZ}^T(M_Z^2)
\ee
and $M_Z\Gamma_Z={\rm Im}\, {\hat\Pi}_{ZZ}^T(k^2)$.  
% Representative superpartner contributions to ${\hat\Pi}_{ZZ}$ are shown in Fig. \ref{fig:nccorr}. 

%\begin{figure}[ht]
%\begin{center}
%\resizebox{6 in}{!}{
%\includegraphics*[20,160][620,620]{nccorrections.ps}}
%\caption{Representative supersymmetric corrections to neutral current observables: (a) $Z$-boson propagator and $Z$-$\gamma$ mixing conributions; (b) vertex corrections; (c) external leg corrections; and (d) box graph contributions}
%\label{fig:nccorr}
%\end{center}
%\end{figure}

The quantities ${\hat g}_V^f$ and ${\hat g}_A^f$ denote the renormalized $Z^0$-fermion vector and axial vector couplings, respectively. While their tree-level values depend on the third component of weak isospin ($I_3^f$) and electric charge ($Q_f$) of fermion $f$ as well as the weak mixing angle, renormalization introduces an additional dependence on a universal renormalization factor ${\hat\kappa}$, along with process-specific vector and axial vector radiative corrections:
\bea
{\hat g}_V^f & = & 2 I_3^f- 4 {\hat\kappa}(k^2,\mu) \sin^2{\hat\theta}_W(\mu)Q_f + {\hat\lambda}_V^f\\
{\hat g}_A^f & = & -2 I_3^f +{\hat\lambda}_A^f \ \ \ .
\eea
Here,  $\sin^2{\hat\theta}_W(\mu)\equiv{\hat s}^2(\mu)$ denotes the weak mixing angle in the ${\overline {\rm MS}}$ scheme:
\be
\label{eq:sinthetadef}
\sin^2{\hat\theta}_W(\mu) = \frac{{\hat g}^\prime(\mu)^2}{{\hat g}(\mu)^2 +{\hat g}^\prime(\mu)^2} \ \ \ ;
\ee
and ${\hat\lambda}_{V,A}^f$  are process-dependent corrections that vanish at tree-level.
Here $\hat{g}$ and $\hat{g}^\prime$ are the ${\rm SU}(2)_L$ and ${\rm U}(1)_Y$ coupling, respectively. 

We emphasize that Eq.~(\ref{eq:sinthetadef}) constitutes a definition of the weak mixing angle in the ${\overline {\rm MS}}$ scheme and it differs in general from the definition in other schemes. For example, on-shell renormalization (OSR) promotes the tree-level relation $\sin^2\theta_W=1-M_W^2/M_Z^2$ to a definition that holds after renormalization. While the OSR and ${\overline {\rm MS}}$ definitions are identical at tree-level, their equality is broken at the one-loop level. Although OSR is often considered more intuitive, it is less conducive for making the most precise SM predictions. In particular,  $\sin^2{\hat\theta}_W(\mu)\equiv{\hat s}^2(M_Z)$ can be obtained from the values of $G_\mu$, $\alpha$, and $M_Z$, which are all known with better precision than $M_W$ as is needed for the OSR definition. Using
\bea
{\hat e}^2(\mu) & =&  {\hat g}^2(\mu) {\hat s}^2(\mu) \\
{\hat M}_W^2 & = & {\hat M}_Z^2 {\hat c}^2 \ \ \ ,
\eea
writing 
\be
{\hat \alpha}(\mu) = \alpha + \Delta{\hat \alpha}(\mu)
\ee
where $\alpha$ is the fine structure constant, employing Eqs.~(\ref{eq:ghat},\ref{eq:mzhat}), and choosing
$\mu=M_Z$ we obtain
\be
\label{eq:Gfswmz}
{\hat s}^2 (M_Z) {\hat c}^2 (M_Z) = \frac{\pi\alpha}{\sqrt{2} M_Z^2 G_\mu\left[1-\Delta{\hat r}(M_Z)\right]}
\ee
where
\be
\label{eq:deltarhat}
\Delta{\hat r}(\mu) = \Delta{\hat r}_\mu+\frac{\Delta{\hat\alpha}}{\alpha} -\frac{{\hat\Pi}_{ZZ}^T(M_Z^2, \mu)}{M_Z^2}\ \ \ .
\ee
Looking ahead to the NC review~\cite{neutral-current}, Eqs,~(\ref{eq:Gfswmz},\ref{eq:deltarhat}) correspond to Eq. (35) in that discussion. There one also finds the related  quantities  $\Delta r$ for OSR and  $\Delta {\hat r}_W$ that is appropriate when using $M_W^2$ in place of ${\hat c}^2 (M_Z) M_Z^2$. A particular advantage of the ${\overline {\rm MS}}$ definition of the weak mixing angle is the absence of a quadratic $m_t$-dependence that enters the OSR definition.  

By itself, ${\hat s}^2(\mu)$ is not an observable since it depends on the renormalization scale. One may, however, define an effective weak mixing angle that is $\mu$-independent and that may in principle be isolated experimentally by comparing experiments with different species of fermions:
\be
\label{eq:sweff}
\sin^2{\hat\theta}_W(k^2)^{\rm eff} \equiv {\hat\kappa}(k^2,\mu) \sin^2{\hat\theta}_W(\mu) \ \ \ .
\ee
Here the quantity ${\hat\kappa}(k^2,\mu)$ introduced earlier describes a class of electroweak radiative corrections that is independent of the species of fermion involved in the NC interaction. Contributions to ${\hat\kappa}(k^2,\mu)$  arise primarily from the $Z$-$\gamma$ mixing tensor:
\be
{\hat\Pi}^{\mu\nu}_{Z\gamma}(k^2) = {\hat\Pi}^T_{Z\gamma}(k^2) T^{\mu\nu} +{\hat\Pi}^{L}_{Z\gamma}(k^2) L^{\mu\nu} \ \ \ .
\ee
%Note that in general, the functions ${\hat\Pi}_{Z\gamma}^T(k^2)$ depend on the choice of electroweak gauge parameter, so to arrive at a gauge-independent ${\hat\kappa}(k^2,\mu)$, a prescription for removing the gauge-dependent components of ${\hat\Pi}_{Z\gamma}^T(k^2)$ must be employed\cite{Ferroglia:2003wa}. 
For processes involving $|k^2| \ll M_Z^2$ that are the focus of this Issue, contributions from light fermions to ${\hat\kappa}(k^2,\mu)$ can lead to the presence of large logarithms when one chooses $\mu=M_Z$. The presence of these logarithms can spoil the expected behavior of the perturbation series unless they are summed to all orders.

To illustrate, consider the amplitude for low-energy, parity-violating M\o ller scattering:
\be
\label{eq:moller1}
{\cal M}_{PV}^{ee} = \frac{G_\mu}{2\sqrt{2}} {\hat \rho}_{\rm NC}(0) {\hat g}_V^e{\hat g_A}^e\ {\bar e}\gamma_\mu e\ {\bar e} \gamma^\mu\gamma_5 e \ \ \ 
\ee
with
\be
\label{eq:moller2}
Q_W^e \equiv {\hat \rho}_{\rm NC}(0)\, {\hat g}_V^e {\hat g}_A^e = {\hat \rho}_{\rm NC}(0)\left[-1+4{\hat\kappa}(0,\mu){\hat s}^2(\mu)+{\hat\lambda}_V^f+ {\hat\lambda}_A^f(-1+4{\hat s}^2)\right]+\cdots
\ee
being the \lq\lq weak charge" of the electron and with the $+\cdots$ indicating box diagram contribution and terms of order $(\alpha/4\pi)^2$. At tree-level (${\hat\kappa}\to 1$, ${\hat\lambda}_{V,A}^e\to 0$), the weak charge is suppressed, since ${\hat s}^2$ is numerically close to $1/4$: $Q_W^{e,\ \rm tree}\sim -0.1$. Inclusion of one-loop SM radiative corrections reduce the magnitude of $Q_W^e$ by nearly 40 \%, owing largely to the near cancellation between the first two terms in Eq. (\ref{eq:moller2}) and the presence of large logarithms in ${\hat\kappa}(0,\mu)$ when $\mu$ is chosen to be $M_Z$ as is conventional\cite{Czarnecki:1995fw}. Given these two considerations, one would expect the relative size of two-loop corrections to $Q_W^e$ to be considerably larger than the nominal $\alpha/4\pi$ scale. 

In order to improve the convergence of the SM prediction for $Q_W^e$, one should like to sum the large logarithms to all orders. The use of the running $ \sin^2{\hat\theta}_W(\mu)$ provides a means for doing so. By choosing $\mu\sim Q$ in both ${\hat\kappa}(k^2,\mu)$  and $ \sin^2{\hat\theta}_W(\mu)$, using the requirement that their product is $\mu$-independent as per Eq.~(\ref{eq:sweff}), and solving the RG equations for $ \sin^2{\hat\theta}_W(\mu)$ as in Ref.~\cite{Erler:2004in}, one effectively moves all the large logarithms from ${\hat\kappa}(k^2,\mu)$ into $\sin^2{\hat\theta}_W(\mu)$ and sums them to all orders. The result is a SM prediction for $\sin^2{\hat\theta}_W(k^2)^{\rm eff}$ with substantially smaller truncation error than would be obtained by the naive application of perturbation theory to one-loop order. Moreover, as first emphasized in Ref.~\cite{Czarnecki:1995fw}, the SM prediction for $\sin^2{\hat\theta}_W(k^2)^{\rm eff}$ provides a useful benchmark against which to compare various NC experimental results. Since ${\hat\kappa}(k^2,Q)-1$ has its expected magnitude of order $\alpha/\pi$, it is also reasonable to use $ \sin^2{\hat\theta}_W(Q)$ for this purpose.

A detailed discussion of the SM prediction for the running of ${\hat s}^2(\mu)$ in the ${\overline {\rm MS}}$ will appear in the article on neutral current observables in this Issue (see also Ref.~\cite{Erler:2004in}), and we refer the reader to that article for details. Results from the most recent work are shown in 
%As with the running QED and QCD couplings, ${\hat\alpha}(\mu)$ and ${\hat\alpha}_s(\mu)$, respectively, the running of the weak mixing angle is a prediction of the SM and provides a useful benchmark for precision studies in the NC sector\cite{Czarnecki:1995fw} . A renormalization group-improved SM prediction for  ${\hat s}^2(\mu)$ in the ${\overline {\rm MS}}$ scheme has recently been carried out in Ref.~\cite{Erler:2004in}, where logarithmic contributions of the form $\alpha^n \ln^n(\mu/\mu_0)$ (with $\mu_0$ being a reference scale) have been summed to all orders. Additional subleading contributions of the form $\alpha^{n+1}\ln^n(\mu/\mu_0)$ and $\alpha\alpha_s^{n+k}\ln^n(\mu/\mu_0)$ with $k=0,1,2$ were also included in that analysis, and a refined estimate of the hadronic physics uncertainty associated with light-quark loops at low scales performed (see below). The results are shown in 
Fig. 1 of that article.

%Fig.~\ref{fig:sin2theta}, where the scale $\mu$ has been chosen to be $Q=\sqrt{|k^2|}$ for a process occurring at squared momentum transfer $k^2$. The reference scale has been chosen to be $\mu_0=M_Z$ and the running of ${\hat s}^2(Q)$ normalized to reproduce its value at the $Z^0$-pole : 
%$\sin^2{\hat\theta}_W(M_Z)=0.23122(15)$ [{\bf update}].  The discontinuities in the curve of Fig.~\ref{fig:sin2theta} correspond to particle thresholds, below which a particle of the corresponding mass decouples from the running. The change in sign of the slope at $Q=M_W$ arises from the difference in sign of the gauge boson and fermion contributions to the $\beta$ function for ${\hat s}^2(\mu)$. 
%Note that threshold matching conditions in the ${\overline {\rm DR}}$-scheme will differ from those in the ${\overline {\rm MS}}$ framework due to the differences in continuation of the Clifford algebra into $d=4-2\varepsilon$ dimensions\cite{Antoniadis:1982vr, Langacker:1992rq}. 

Of particular interest for this Issue is the weak mixing angle at $Q=0$, which takes on the value\cite{Erler:2004in}:
\be
\sin^2{\hat\theta}_W(0)= 0.23867\pm 0.00016\ \ \ .
\ee
Note that the error is dominated by the experimental error in $\sin^2{\hat\theta}_W(M_Z)$ and that the value of ${\hat s}^2(\mu)$ at the two scales differs by roughly three percent. Several present and prospective NC experiments are sensitive to this running, including measurements of the PV asymmetries in M\"oller scattering, the PV asymmetries in both elastic and inelastic electron-hadron scattering,  and atomic PV observables that are insensitive to the nuclear spin. The PV M\o ller and elastic $ep$ asymmetries are particularly sensitive to $\sin^2{\hat\theta}_W(0)$, as they depend on $1-4{\hat\kappa}\sin^2{\hat\theta}_W(0)\approx 0.1$. Because of this fortuitous suppression, relatively small changes in $\sin^2{\hat\theta}_W(0)$ (such as the three percent effect due to running) can lead to considerably larger effects in the asymmetries. Consequently, inclusion of the running effect in the weak mixing angle is vital to the interpretation of these low-energy observables in terms of possible BSM physics.

%\begin{figure}
%\begin{center}
%\includegraphics[width=4in, angle=-90]{sin2theta.ps}
%\caption{Calculated running of the weak mixing angle in the SM, defined in the
%$\overline{\rm MS}$ renormalization scheme.
%Also shown are the experimental results from APV,
%neutrino DIS ($\nu$-DIS),   
%parity violating asymmetry measurement at E158 ($A_{PV}$), 
%the expected precision of Qweak 
%and the lepton forward-backward asymmetry measurement
%at CDF ($A_{FB}$).  This plot
%is taken from Ref.\cite{sin2theta}.}
%\label{fig:sin2theta}
%\end{center}
%\end{figure}

%\begin{figure}[ht]
%\begin{center}
%\includegraphics[width=4in]{sin2thetarunning.ps}
%\caption{Scale-dependence of the weak mixing angle, $\sin^2{\hat\theta}_W(\mu)$ in the  ${\overline %{\rm MS}}$ scheme. }
%\label{fig:sin2theta}
%\end{center}
%\end{figure}

%In the case of superpartner loop contributions to low-energy observables, it is sufficient to include their effects solely in the form factor ${\hat\kappa}(k^2,\mu)$ while choosing $\mu=M_Z$ (illustrative contributions to ${\hat\Pi}_{Z\gamma}$ are shown in Fig. \ref{fig:nccorr}). In addition, one should include their effects in the value of $\sin^2{\hat\theta}_W(M_Z)$. One may adopt two different strategies for doing so: 

The corrections contained in the ${\hat\lambda}_{V,A}^f$ contain the $Zff$ vertex and external leg corrections and are specific to the fermion species. A similar remark applies to the box graphs that contribute to the four-fermion amplitudes and that cannot be absorbed into the  individual vector and axial vector couplings.  An additional contribution to the four fermion amplitudes is generated by  $\gamma$ exchange and involves the so-called anapole coupling of the fermion\cite{zeldovich,Musolf:1990sa}
\be
\label{eq:anapole1}
{\cal L}_{\rm anapole} = \frac{eF_A}{M^2} {\bar\psi} \gamma_\mu\gamma_5 \psi \ \partial_\nu F^{\mu\nu} \ \ \ ,
\ee
where $F_A$ is the dimensionless anapole moment and $M$ is an appropriate mass scale. The interaction ${\cal L}_{\rm anapole}$ generates a contribution to the fermion matrix element of the electromagnetic current:
\bea
\label{eq:anapole2}
\bra{p'} J_\mu^{EM}(0)\ket{p} & = & {\bar U}(p') \Bigl[ F_1 \gamma_\mu + \frac{iF_2}{2 M} \sigma_{\mu\nu} k^\nu \\
\nonumber
&& + \frac{F_A}{M^2} \left(k^2\gamma_\mu -\dslash{k}k_\mu\right)\gamma_5 +\frac{iF_E}{2 M}\sigma_{\mu\nu} k^\nu\gamma_5\Bigr] U(p) \ \ \ ,
\eea
where $k=p'-p$ and where $F_1$, $F_2$, $F_A$, and $F_E$ give the Dirac, Pauli, anapole, and electric dipole form factors, respectively\footnote{Note that the overall sign of the anapole term in Eq.~(\ref{eq:anapole2}) differs from the convention used in Ref.~\cite{Kurylov:2003zh}.}.  Since only weak interactions can give rise to the  parity-odd photon-fermion anapole  coupling, we choose $M=M_Z$ in Eq.~(\ref{eq:anapole1}). 
From  Eq. (\ref{eq:anapole1}) one sees that the anapole coupling gives rise to a contact interaction in co-ordinate space, since $\partial_\nu F^{\mu\nu} = j_\mu$ with $j_\mu$ being the current of the other fermion involved in the low-energy interaction. 

We note that the coupling $F_A$ itself depends on the choice of electroweak gauge, and only the complete one-loop scattering amplitude that includes all ${\cal O}(\alpha)$ electroweak radiative corrections (including $F_A$) is gauge-independent (see Ref.~\cite{Musolf:1990sa} and references therein). Nonetheless, when classifying various topologies of the one-loop corrections, it is useful to separate out the anapole contributions that behave like an effective contribution to the  product of vector coupling ${\hat g}_V^f$ and the axial vector coupling ${\hat g}_A^{f^\prime}$:
\be
\left( {\hat g}_A^{f^\prime} {\hat g}_V^f \right)_{\rm anapole} = -16{\hat c}^2{\hat s}^2 Q_f F_A^{f^\prime}\ \ \ .
\ee

Adding the anapole contribution to those from the other one-loop corrections leads to a gauge-invariant scattering amplitude. Analogous $\gamma$-exchange effects  enter in the scattering amplitudes for neutrino scattering from charged particles. In this case, the anapole contribution is equivalent to the neutrino charge radius. In the NC review, the charge radii/anapole, box graph, and vertex plus external leg corrections, and gauge boson propagator corrections are separately discussed, though one should keep in mind this classification carries a gauge-dependence.

%\begin{figure}[ht]
%\begin{center}
%\resizebox{5 in}{!}{
%\includegraphics*[60,520][470,640]{anapole.ps}}
%\caption{Anapole contributions to the NC interaction between two fermions.}
%\label{fig:anapole}
%\end{center}
%\end{figure}

When looking beyond the SM radiative corrections to  possible loop-induced BSM contributions to low-energy precision NC observables, it is important to include constraints from higher-energy studies. Prior to the operation of the LHC, the most important constraints had been obtained from  precision $Z$-pole observables. In the corresponding theoretical interpretation, it has been useful to characterize possible corrections to the gauge boson propagators from new heavy particles 
in terms of the so-called oblique parameters, $S$, $T$, $U$ \cite{Peskin:1990zt,Golden:1990ig,Marciano:1990dp, Kennedy:1990ib,Kennedy:1991sn,Altarelli:1990zd,Holdom:1990tc,Hagiwara:1994pw}: 

\begin{eqnarray}
\label{eq:stu-sirlin}
S&=&\frac{4{\hat s}^2{\hat c}^2}{{\hat \alpha}M_Z^2}{\rm Re}\Biggl\{
{\hat \Pi}_{ZZ}(0)-{\hat \Pi}_{ZZ}(M_Z^2)+\frac{{\hat c}^2-{\hat
s}^2}{{\hat c}{\hat s}} \left[{\hat \Pi}_{Z\gamma}(M_Z^2)-{\hat
\Pi}_{Z\gamma}(0)\right] +{\hat \Pi}_{\gamma\gamma}(M_Z^2)
\Biggr\}^{\rm New} ~,\nonumber \\ 
T&=&\frac{1}{{\hat \alpha}M_W^2}
\Biggl\{ 
{\hat c}^2\left( {\hat \Pi}_{ZZ}(0)+\frac{2{\hat s}}{\hat c}
{\hat \Pi}_{Z\gamma}(0) \right) -{\hat \Pi}_{WW}(0) \Biggr\}^{\rm
New} ~,\nonumber \\ 
U&=&\frac{4{\hat s}^2}{\hat \alpha} \Biggl\{
\frac{{\hat \Pi}_{WW}(0)-{\hat \Pi}_{WW}(M_W^2)}{M_W^2} +{\hat
c}^2\frac{{\hat \Pi}_{ZZ}(M_Z^2)-{\hat \Pi}_{ZZ}(0)}{M_Z^2} \nonumber
\\ 
&+&2{\hat c}{\hat s}
\frac{ {\hat \Pi}_{Z\gamma}(M_Z^2)-{\hat
\Pi}_{Z\gamma}(0)}{M_Z^2} +{\hat s}^2 \frac{{\hat
\Pi}_{\gamma\gamma}(M_Z^2)}{M_Z^2} \Biggr\}^{\rm New}
~,\end{eqnarray}
where the superscript \lq\lq New" indicates that only the new physics
contributions to the self-energies are included.  Contributions to
gauge-boson self energies  can be expressed entirely in terms of the
oblique parameters $S$, $T$, and $U$ in the limit that $\Lambda\gg
\mz$. 

%However, since present collider limits allow for fairly light
%superpartners, we do not work in this limit\footnote{It is possible to extend the oblique parameterization in this case with three additional parameters\cite{Maksymyk:1993zm}. For the low-energy observables of interest here, this extended oblique approximation is not especially useful.}. Consequently, the
%corrections arising from the photon self-energy ($\Pi_{\gamma\gamma}$)
%and $\gamma$-$Z$ mixing tensor ($\Pi_{Z\gamma}$) contain a residual
%$k^2$-dependence not embodied by the oblique parameters.  

Expressing  BSM loop contributions to 
${\hat\rho}$ and $\sin^2{\hat\theta}_W(k^2)^{\rm eff} = {\hat\kappa}(k^2,\mu) \sin^2{\hat\theta}_W(\mu)$ in terms of $S$,$T$, and $U$ we obtain:
\begin{eqnarray}
\delta{\hat\rho}^{\rm BSM\, loop} & = & {\hat\alpha} T-{\hat\delta}_{VB}^\mu
~,\nonumber \\ 
\nonumber \\
\left(\frac{\delta\sin^2{\hat\theta}_W^{\rm eff}}{\sin^2{\hat\theta}_W^{\rm eff}}\right)^{\rm BSM\, loop} & = & \left(
\frac{{\hat c}^2}{{\hat c}^2-{\hat s}^2} \right) 
\left(\frac{{\hat\alpha}}{4{\hat s}^2
{\hat c}^2} S-{\hat \alpha} T +{\hat\delta}_{VB}^\mu \right) +
\frac{{\hat c}}{{\hat s}}\Bigl[  \frac{{\hat\Pi}_{Z\gamma}(k^2)}{k^2}-
\frac{{\hat\Pi}_{Z\gamma}(M_Z^2)}{M_Z^2}\Bigr] \nonumber \\
&&+\Bigl(\frac{{\hat c}^2}{{\hat c}^2-{\hat s}^2}
\Bigr)\Bigl[-\frac{{\hat\Pi}_{\gamma\gamma}(M_Z^2)}{M_Z^2}
+\frac{\Delta{\hat\alpha}}{\alpha} \Bigr] ~,
\label{eq:rho-kappa-stu}
\end{eqnarray}
where
$k^2$ is the typical momentum
transfer for a given process.  For low energy interactions, 
$k^2\rightarrow 0$. Note that we have included in  $\delta\sin^2{\hat\theta}_W^{\rm eff}$ both the the contribution from ${\hat\Pi}_{Z\gamma}(k^2)/k^2$ that enters ${\hat\kappa}(k^2,\mu)$ as well as the shift in ${\hat s}^2(M_Z^2)$ obtained from Eq.~(\ref{eq:Gfswmz}) as discussed above. 
Eqs.~(\ref{eq:rho-kappa-stu}) provide a useful means of incorporating $Z^0$-pole constraints on BSM loop effects. For example, $\delta{\hat\rho}^{\rm BSM\, loop}$ is highly constrained by bounds on $T$ obtained from such observables. In contrast, $[\delta \sin^2{\hat\theta}_W(k^2)^{\rm eff}]^{\rm BSM\, loop}$ is less stringently constrained. 

%As we discuss in Section \ref{sec:nc} below, the  unconstrained contributions to the effective weak mixing angle can lead to relatively large effects in some low-energy NC processes.

\subsection{\it Theoretical Uncertainties in Electroweak Radiative Corrections}

An important consideration in exploiting low-energy, precision electroweak observables as a probe  of BSM physics is to ensure that the theoretical uncertainties associated with SM contributions are well-below the level of possible BSM effects. The SM uncertainties generally involve one of two considerations: (i) neglect of higher order electroweak contributions, and (ii) contributions from strong interactions. We have already touched on the latter in our short discussion of $g_\mu-2$. While an extensive discussion of these considerations goes beyond the scope of the present article, we give here a brief overview of the strategies employed to address them. 

Nominally, one expects the one-loop contributions to quantities such as ${\Delta \hat{r}}_\mu$, $\hat\kappa$, {\em etc.} to be of order $\alpha/\pi\sim 10^{-3}$, so that neglect of two- and higher-loop effects is well justified for the present level of experimental sensitivity. Moreover, since SUSY loop contributions must generally decouple in the ${\Lbsm}\to\infty$ limit, one expects the relative magnitude of their contributions to be
\be
\delta_{\rm BSM\ loop} = \frac{{\delta\cal O}^{\rm BSM\ loop}}{{\cal O}^{\rm SM}} \sim \frac{\alpha}{\pi}\left(\frac{M}{\Lbsm}\right)^2 \  \  \ ,
\ee
where $M$ is the relevant SM mass. For low-energy electroweak processes, one has $M\to v$, so that for $\Lbsm\gsim v$  one  expects $\delta_{\rm BSM\ loop}$ to be somewht smaller than, the scale of one-loop, SM electroweak corrections. From this standpoint, neglect of two-loop SM contributions is a justifiable approximation. As discussed above, however,  exceptions may occur when the one-loop SM contributions contain large logarithms (as in the case of ${\hat\kappa}$), when the tree-level SM amplitudes are suppressed ({\em e.g.}, NC amplitudes proportional to ${\hat g}_V^e$), or both. In such situations, the summing terms of the form $\alpha^n \ln^n(\mu/\mu_0)$ is essential, and the RG equations can be employed for this purpose. 

Reduction of theoretical uncertainties associated with QCD corrections is generally more challenging. Short-distance QCD contributions can  be treated using the operator product expansion (OPE), and the resulting correction to a given order in $\alpha_s$ achieved with sufficient effort. In the case of PV electron-proton scattering, for example, the one-loop $WW$ box contribution is anomalously -- but not logarithmically -- enhanced, and its contribution to the proton weak charge, $Q_W^p$, nearly cancels that of the large logarithms appearing in ${\hat\kappa}$ [or resummed into $\sin^2{\hat\theta}_W(0)$]. Since the semileptonic, $WW$ box graphs involve hadronic intermediate states one could expect relatively important QCD corrections to the one-loop amplitude. Because the loop integral is  dominated by high momentum scales  the corrections can be computed using the OPE, leading to \cite{Erler:2003yk}
\be
\delta Q_W^p(WW\ {\rm box}) = \frac{\hat\alpha}{4\pi{\hat s}^2}\left[2+5\left(1-\frac{\alpha_s(M_W)}{\pi}\right)\right]\ \ \ 
\ee
for a total QCD correction of $\approx -0.7\%$. 

A more problematic situation arises for one-loop corrections that sample momenta of order the hadronic scale.  To illustrate, we first consider PV electron scattering. For both PV M\o ller and elastic $ep$ scattering, light quark loop contributions to ${\hat\Pi}_{Z\gamma}^T$ lead to hadronic uncertainties in ${\hat\kappa}(0,\mu)$. Traditionally, light quark contributions have been computed by relating ${\hat\Pi}_{Z\gamma}^T$ to the $\sigma(e^+ e^-\to{\rm hadrons})$ via dispersion relations\cite{Marciano:1982mm,Marciano:1983ss}, much as one does in computing hadronic vacuum polarization contributions to $a_\mu$. In the case of ${\hat\Pi}_{Z\gamma}^T$, however, additional assumptions regarding flavor symmetry in the current-current correlator are needed in order to make use of $e^+ e^-$ data. Recently, these assumptions have been examined and more stringent bounds on the hadronic uncertainty in ${\hat\kappa}(0,\mu)$ obtained\cite{Erler:2004in}. 

For semileptonic processes, additional hadronic uncertainties appear in box graphs that contain one $\gamma$ and one weak gauge boson. In contrast to the situation for the $WW$-box graphs, the $\gamma Z$ loop integral samples all momentum from the hadronic scale to the weak scale. Neglecting the short-distance, perturbative QCD corrections, one finds
\be
\label{eq:zgbox}
\delta Q_W^p(\gamma Z\ {\rm box}) = \frac{5\hat\alpha}{2\pi}\left(1-4{\hat s}^2\right)\left[\ln\left(\frac{M_Z^2}{\Lambda^2_H}\right)+C_{\gamma Z}(\Lambda_H)\right]\ \ \ ,
\ee
where the hadronic scale $\Lambda_H$ is a scale characterizing the transition between the perturbative and non-perturbative domains\footnote{This scale is denoted elsewhere in this Issues as the scale of chiral symmetry-breaking, $\Lambda_\chi$.} and $C_{\gamma Z}(\Lambda_H)$ parameterizes contributions to the loop integral from momenta
$\sqrt{|k^2|} \lsim \Lambda_H$. The coefficient of  logarithm in Eq. (\ref{eq:zgbox}) is determined by short distance dynamics and can be calculated reliably in perturbation theory. However, the value of  $C_{\gamma Z}(\Lambda_H)$ is sensitive to long-distance scales and has not, as yet, been computed from first principles in QCD. A similar contribution arises in neutron, nuclear, and pion $\beta$-decay. An estimate of the theoretical uncertainty associated with these contributions had been made by varying $\Lambda$ over the range $ 400 \leq \Lambda_H \leq 1600$ MeV. The corresponding variation in the logarithmic term in Eq.~(\ref{eq:zgbox}) was used as an indication of the uncertainty associated with long-distance contributions to the box graph integral. Recently, Marciano and Sirlin observed that for the $\gamma W$ box, both the pQCD corrections to the logarithmic term as well as the value of $\Lambda_H$ could be obtained by comparison with the theoretical expression for the Bjorken Sum Rule using isospin symmetry\cite{Marciano:2005ec}. As a result, these authors have reduced the previously-quoted theoretical error by a factor of two. The analogous treatment of the $\gamma Z$ box is more complex, since one cannot obtain the isoscalar contribution from isospin arguments. In both cases, the more refined estimates of the  uncertainty associated with the low-energy constants $C_{\gamma Z}$ and $C_{\gamma W}$ remain to be performed.

For PV electron scattering, there exists an additional contribution to the PV asymmetry associated with the energy-dependence of the $\gamma Z$ box graphs. As an energy-dependent effect, this contribution is not formally part of the fundamental, renormalized electroweak couplings, but rather more akin to a form factor. Nonetheless, the contribution can introduce an additional source of theoretical uncertainty into the extraction of the fundamental couplings from the asymmetry. Consequently, reducing this uncertainty constitutes one of the on-going theoretical challenges for the interpretation of the PV electron scattering experiments. A detailed discussion of this correction and its quantitive impact on the interpretation of the asymmetry measurements can be found in the NC article in this issue as well as in Ref.~\cite{Kumar:2013yoa}.

\section{Beyond the Standard Model}
\label{sec:BSM}

%%%%%%%%%%%%%%%%%%%%%%%%%%%%%%%%%%%%%%%%%%%%%%%%%%%%%%%%
%        FIGURE
%%%%%%%%%%%%%%%%%%%%%%%%%%%%%%%%%%%%%%%%%%%%%%%%%%%%%%%%
\begin{figure}[t]
\centering 
\includegraphics[width=0.8\textwidth]{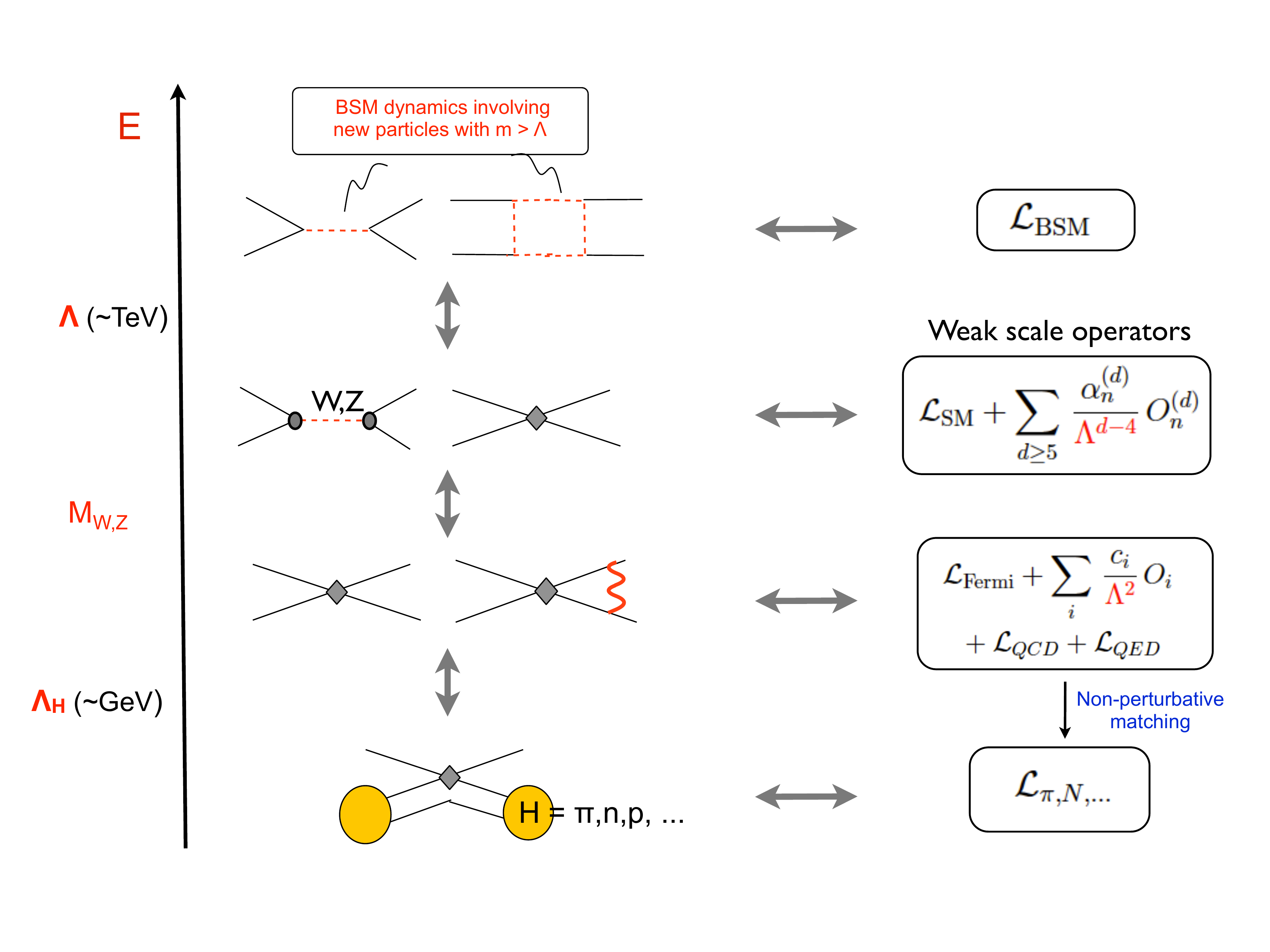}
\begin{minipage}[t]{16.5 cm}
\caption{Schematic representation of the  series of effective field theories (EFTs) 
needed to describe the influence of new physics beyond the Standard Model 
on  low-energy observables (see text for details).\label{fig:EFT}}
\end{minipage}
\end{figure}

\subsection{\it  Effective Theory  Description: generalities  \label{sec:BSMEFT}}

While the SM successfully describes a wealth of data over a wide range of 
energies (from atomic scales to hundreds of GeV),  both purely theoretical arguments 
and observed facts about our Universe point to the existence of new 
degrees of freedom and interactions beyond the SM. 
With the theoretical bias that new physics originates at high scales
(or short distances),  we can  think of the SM as the low-energy limit of  a more  fundamental theory. 
~\footnote{Strictly speaking, this way of thinking does not apply to 
SM extensions  that involve light new particles very weakly coupled to the SM.} 

In this context and  in the absence of an emerging ``New Standard Model", it is convenient to 
describe the dynamics  below the scale $\Lambda$ 
(at which new particles appear)  through an effective field theory (EFT), in which 
the new  heavy particles are ``integrated out" and  affect the dynamics through 
a series of higher dimensional  operators constructed with the low-energy SM fields. 
The basic ideas of  this approach are illustrated  in Fig.~\ref{fig:EFT}, which shows the relevant scales involved in the problem 
and how short-distance physics beyond the SM works its way into low-energy probes: 

\begin{itemize}
\item  Above the  scale $\Lambda$, where the new particles live, the dynamics is described by the full UV completion of the 
SM,  characterized by a Lagrangian density ${\cal L}_{\rm BSM}$. 

\item Below  the  scale $\Lambda$, 
one can ``integrate out" the new degrees of freedom. 
The effective Lagrangian relevant at scales $\Lambda > E >   M_{Z,W}$, 
is the SM Lagrangian augmented by a string of 
$d>4$  operators constructed with the low-energy SM fields, suppressed by $\Lambda^{d-4}$.
Any UV model analysis can be cast in this language through a matching calculation at the scale $\Lambda$,   
which relates  the  effective couplings  to the couplings and masses of  the model. 
The prototypical example of such a matching calculation is the relation between the Fermi constant and  the $W$ boson mass and 
$SU(2)$ gauge coupling of the SM.

\item Below the electroweak scale one integrates out $W$ and $Z$,   and the dynamics
at $ M_{Z,W} >  E >  \Lambda_{H}$  (where $\Lambda_H \sim 1$~GeV)  
is described by a  modified set of effective operators (BSM plus electroweak), plus QCD and QED. 
This picture still involves  quarks and gluons as explicit degrees of freedom.

\item Finally,  to describe  processes involving hadrons and nuclei at $E \leq  \Lambda_H$, 
one has to go from a picture of quarks and gluons  to a picture of 
free or bound hadrons.  This requires non-perturbative matching calculations.
The step from quarks to hadrons exploits the symmetries of QCD and when symmetries 
are not sufficient   typically requires a combination of  lattice QCD and Chiral Perturbation Theory (ChPT). 
The description of nuclei requires the use of  non-perturbative  non-relativistic many-body techniques.

\end{itemize}

% 1
From the above quick overview both the benefits and limitations of the EFT description should be clear. 
First,  the EFT approach   is quite general and allows one  to study the implications
of low-energy measurements  on a large class of  models.
% 2 
In particular, note that the operator analysis is certainly applicable  to describe 
the low-energy probes we are mostly concerned in this  review. 
This method enables us to 
assess in a model-independent way  the  possible correlations, relative significance,  and impact 
of various  low-energy  probes  
(i.e. those observables sensitive to the same set of BSM effective couplings and operators). 
% 3 
Moreover, if the new physics scale  $\Lambda$  is in the multi-TeV region or above 
(which is  an open question),  even at LHC energy scales $E \sim 1$~TeV  the BSM dynamics 
can be described in terms of effective operators.  
In other words,   if new physics originates at very high energy ($E > {\rm few}$~TeV), the operator 
analysis can be used to directly compare low-energy  versus   collider  probes.
%4

Explicit model analyses are related to the EFT description by matching calculations 
at the scale $\Lambda$. The matching conditions express the effective couplings in terms of the fewer 
model couplings and masses, implying specific model correlations among the various low-energy 
couplings (and observables) and the collider phenomenology. 
In general such correlations are not ``visible" in a pure bottom-up EFT approach. 
In this set of reviews we will use both a general EFT description of low-energy probes
and explicit UV models.

%\subsection{\it  Weak scale effective lagrangian
\subsection{\it  BSM effective lagrangian
\label{sect:weakscale}}

In this subsection we provide  some details on the 
%weak-scale effective Lagrangian. 
BSM effective Lagrangian at the TeV scale. 
This  BSM effective Lagrangian provides the starting point for low-energy analyses of various types of probes 
(CP violation and EDMs,   lepton flavor violation,   non-standard CC weak interactions, etc), 
which will be discussed in  each individual chapter. 

As discussed earlier, given the successes of the SM
at energies up to the  electroweak  scale  $v \sim 200$ GeV,
we adopt here  the point of view that
the SM is the low-energy limit of a more fundamental theory. 
Writing down the effective theory requires specifying (i) a power counting in ratio of scales; 
(ii) the low-energy degrees of freedom (field content); and  (iii) the symmetries respected by the 
underlying physics.   Each of these points comes with some assumptions about the 
underlying dynamics, which we now briefly discuss:
\begin{itemize}
\item Power counting: we assume that  there is a  gap between the weak scale $v$
and the scale $\Lambda$  where new degrees of freedom appear, We organize the 
expansion in dimensionality of the new physics operators, whose coefficients scale 
with inverse powers of $\Lambda$, namely $1/\Lambda^{d-4}$.  
To a given order in $v/\Lambda$ the EFT is ``renormalizable", in the sense 
that all the ultraviolet divergences can be reabsorbed in a finite number of parameters.
\item Degrees of freedom: we need to specify the building blocks of the effective theory. 
Our default assumptions will be that  the low-energy field content is the same as in the SM. 
Overall, this is  a safe assumption, except for two cases, in which this becomes a 
model assumption and deserves further discussion. 
(1)  Neutrino field content:  by making the SM field choice,  we
assume that there are no  ligh  right-handed neutrino fields, sterile with respect to the SM gauge group.
This  excludes a Dirac-type mass term for neutrinos but  at the moment there is no experimental evidence 
to support this choice.  So in some applications to be discussed below 
we will extend the field content to include  $\nu_R$ in the low-energy theory. 
(2) By including the SM Higgs doublet in the low-energy theory, we are making a strong 
assumption about the mechanism of EWSB. 
While we know that EW symmetry is broken and the Higgs mechanism is at work (the Goldstone modes 
become the longitudinal components of the massive gauge bosons), 
we do not  conclusively know how this is realized. 
By including the SM Higgs in the low-energy theory, 
we assume that the EWSB 
sector is weakly coupled  and EWSB proceeds as in the SM. 
This scenario is receiving growing support by the LHC data on the Higgs.  
The possibility of  a Higgs-less effective theory, corresponding to 
a strongly interacting EWSB sector,  looks less likely right now and for simplicity we will not discuss 
it in detail  in this review.   
EFT  analyses in this context can be found in~\cite{Appelquist:1993ka, Longhitano:1980iz, Feruglio:1992wf,Wudka:1994ny}.
%Similarly,  analyses of certain sectors of the theory involving light right-handed neutrinos can be found 
%in  Refs.~{\bf [REFS]}. 

\item Symmetries: one needs to specify the symmetries obeyed by the underlying 
theory. Here again we will make the standard assumptions of Lorentz invariance and 
invariance under the gauge SM  group (as part of  a possibly broader set of underlying symmetries). 
So the operators induced by the underlying theory will be Lorentz-invariant structures symmetric 
under  SU(3)$_C\times$SU(2)$_L\times$U(1)$_Y$. 
In particular, we do not impose any of the global ``accidental" symmetries of the SM,  
such as baryon number, lepton number, and lepton flavor.~\footnote{These symmetries 
are not imposed on the SM Lagrangian: 
but it turns out that all the gauge-invariant operators of dimension  less than or equal to four 
respect those symmetries, hence the characterization ``accidental".}

 \end{itemize}
%

%In order to proceed and write down the EFT Lagrangian, one has to make a number of 
%technical assumptions, namely that: 
%(i) there is a  gap between the weak scale $v$
%and the scale $\Lambda$  where new degrees of freedom appear;
%(ii)
% the SM extension at the weak scale is  weakly coupled,
%so the EW symmetry $SU(2)_L \times U(1)_Y$ is linearly realized
%and the low-energy theory contains a SM-like  Higgs doublet~\cite{Buchmuller:1985jz}. {\bf [CLARIFY / SIMPLIFY]}
%(iii) 
%The low-energy field content is the same as in the SM, i.e. there are no ``light"
% right-handed neutrino fields (sterile with respect to the 
%SM gauge group).
%% 

Given the above discussion, one can 
describe physics at the weak scale (and below) by means of an
effective non-renormalizable lagrangian  of the form:
\bea
\label{eq:EFT}
%{\cal L}^{(\rm{eff})} &=& {\cal L}_{\rm{SM}} + \frac{1}{\Lambda} {\cal L}_5 + \frac{1}{\Lambda^2} {\cal L}_6 +   \ldots \\
%{\cal L}_n &=& \sum_i \alpha^{(n)}_i~ Q_i^{(n)}~,
{\cal L}^{(\rm{eff})} &=& {\cal L}_{\rm{SM}} +  \ {\cal L}_5 +  {\cal L}_6 +   \ldots \\
{\cal L}_d &=& \sum_i       \frac{\alpha_i^{(d)}}{\Lambda^{d-4}}   ~ Q_i^{(d)}~,
\eea
where $\Lambda $ is the scale of the new physics associated with local $SU(2)_L \times U(1)_Y$  gauge-invariant operators $Q_i^{(n)}$ of dimension $d$ built out of SM fields and where the Wilson coefficients $\alpha_i^{(d)}$ embody the details of the underlying dynamics at the energy scale $\Lambda$. For example, $\Lambda$ may denote the mass scale of new degrees of freedom that become active at the BSM energy scale, while the $\alpha_i^{(d)}$ may contain loop factors, couplings, and/or other factors associated with a particular symmetry violation such as CP-violating phases. 

Experiments at the intensity / precision frontier  typically probe the 
scale associated with the new operators and their symmetry structure. 
The operators themselves can be divided in two classes: (i)~those that produce 
corrections to SM allowed processes, and can thus be probed via 
precision tests; (ii) those that violate exact or approximate SM symmetries and hence 
mediate forbidden or rare processes. 
Below, we organize our discussion by increasing dimensionality of the operators.

\subsection{\it  Dimension five
\label{sect:dim5}}

At dimension five, only one operator arises~\cite{Weinberg:1979sa}. 
It violates total lepton number and after EWSB generates a Majorana mass term for neutrinos:
\be
\label{eq:mnumajorana}
\mathcal{L}^\nu_\mathrm{Majorana} = - {\tilde Y}_\nu^{jk} \frac{1}{\Lambda}\left({\ell}_j \right)^T   C \,  \epsilon \varphi \, \varphi\epsilon \ell_k~. 
\ee
For $v=246$ GeV, the couplings $\tilde{Y}_\nu$  in  Eq.~(\ref{eq:mnumajorana}) can be $\mathcal{O}(1)$ if $\Lambda\gsim 10^{14}$ GeV. The well-known Seesaw mechanism for neutrino mass generates the Majorana mass term by introducing additional heavy Majorana neutrino fields $N_R$ that couple to the left-handed SM doublets $\ell$ through interactions of the type (\ref{eq:mnudirac}) with the replacement $\nu_R\to N_R$. When one integrates the $N_R$ out of the low energy effective theory, one obtains (\ref{eq:mnumajorana}) with $\Lambda=m_{N_R}$. While the $N_R$ could not be directly observed due to their large mass, the existence of the Majorana mass term could provide indirect evidence for their existence. As discussed in detail in the article on lepton flavor and number violation, the observation of the lepton-number violating $0\nu\beta\beta$ process would imply the presence of non-vanishing Majorana couplings ${\tilde Y}_\nu$. 

It is remarkable that  the first evidence for BSM physics (neutrino mass) can be accounted by the  lowest-dimensional 
effective operator that one can write down.

\begin{table}[t] 
\centering
\renewcommand{\arraystretch}{1.5}
\btb{||c|c||c|c||c|c||} 
\hline \hline
\multicolumn{2}{||c||}{$X^3$} & 
\multicolumn{2}{|c||}{$\vp^6$~ and~ $\vp^4 D^2$} &
\multicolumn{2}{|c||}{$\psi^2\vp^3$}\\
\hline
$Q_G$                & $f^{ABC} G_\mu^{A\nu} G_\nu^{B\rho} G_\rho^{C\mu} $ &  
$Q_\vp$       & $(\vp^\dag\vp)^3$ &
$Q_{e\vp}$           & $(\vp^\dag \vp)(\bar l_p e_r \vp)$\\
$Q_{\wt G}$          & $f^{ABC} \wt G_\mu^{A\nu} G_\nu^{B\rho} G_\rho^{C\mu} $ &   
$Q_{\vp\Box}$ & $(\vp^\dag \vp)\raisebox{-.5mm}{$\Box$}(\vp^\dag \vp)$ &
$Q_{u\vp}$           & $(\vp^\dag \vp)(\bar q_p u_r \tvp)$\\
$Q_W$                & $\eps^{IJK} W_\mu^{I\nu} W_\nu^{J\rho} W_\rho^{K\mu}$ &    
$Q_{\vp D}$   & $\left(\vp^\dag D^\mu\vp\right)^\star \left(\vp^\dag D_\mu\vp\right)$ &
$Q_{d\vp}$           & $(\vp^\dag \vp)(\bar q_p d_r \vp)$\\
$Q_{\wt W}$          & $\eps^{IJK} \wt W_\mu^{I\nu} W_\nu^{J\rho} W_\rho^{K\mu}$ &&&&\\    
\hline \hline
\multicolumn{2}{||c||}{$X^2\vp^2$} &
\multicolumn{2}{|c||}{$\psi^2 X\vp$} &
\multicolumn{2}{|c||}{$\psi^2\vp^2 D$}\\ 
\hline
$Q_{\vp G}$     & $\vp^\dag \vp\, G^A_{\mu\nu} G^{A\mu\nu}$ & 
$Q_{eW}$               & $(\bar l_p \sigma^{\mu\nu} e_r) \tau^I \vp W_{\mu\nu}^I$ &
$Q_{\vp l}^{(1)}$      & $(\vpj)(\bar l_p \gamma^\mu l_r)$\\
$Q_{\vp\wt G}$         & $\vp^\dag \vp\, \wt G^A_{\mu\nu} G^{A\mu\nu}$ &  
$Q_{eB}$        & $(\bar l_p \sigma^{\mu\nu} e_r) \vp B_{\mu\nu}$ &
$Q_{\vp l}^{(3)}$      & $(\vpjt)(\bar l_p \tau^I \gamma^\mu l_r)$\\
$Q_{\vp W}$     & $\vp^\dag \vp\, W^I_{\mu\nu} W^{I\mu\nu}$ & 
$Q_{uG}$        & $(\bar q_p \sigma^{\mu\nu} T^A u_r) \tvp\, G_{\mu\nu}^A$ &
$Q_{\vp e}$            & $(\vpj)(\bar e_p \gamma^\mu e_r)$\\
$Q_{\vp\wt W}$         & $\vp^\dag \vp\, \wt W^I_{\mu\nu} W^{I\mu\nu}$ &
$Q_{uW}$               & $(\bar q_p \sigma^{\mu\nu} u_r) \tau^I \tvp\, W_{\mu\nu}^I$ &
$Q_{\vp q}^{(1)}$      & $(\vpj)(\bar q_p \gamma^\mu q_r)$\\
$Q_{\vp B}$     & $ \vp^\dag \vp\, B_{\mu\nu} B^{\mu\nu}$ &
$Q_{uB}$        & $(\bar q_p \sigma^{\mu\nu} u_r) \tvp\, B_{\mu\nu}$&
$Q_{\vp q}^{(3)}$      & $(\vpjt)(\bar q_p \tau^I \gamma^\mu q_r)$\\
$Q_{\vp\wt B}$         & $\vp^\dag \vp\, \wt B_{\mu\nu} B^{\mu\nu}$ &
$Q_{dG}$        & $(\bar q_p \sigma^{\mu\nu} T^A d_r) \vp\, G_{\mu\nu}^A$ & 
$Q_{\vp u}$            & $(\vpj)(\bar u_p \gamma^\mu u_r)$\\
$Q_{\vp WB}$     & $ \vp^\dag \tau^I \vp\, W^I_{\mu\nu} B^{\mu\nu}$ &
$Q_{dW}$               & $(\bar q_p \sigma^{\mu\nu} d_r) \tau^I \vp\, W_{\mu\nu}^I$ &
$Q_{\vp d}$            & $(\vpj)(\bar d_p \gamma^\mu d_r)$\\
$Q_{\vp\wt WB}$ & $\vp^\dag \tau^I \vp\, \wt W^I_{\mu\nu} B^{\mu\nu}$ &
$Q_{dB}$        & $(\bar q_p \sigma^{\mu\nu} d_r) \vp\, B_{\mu\nu}$ &
$Q_{\vp u d}$   & $i(\tvp^\dag D_\mu \vp)(\bar u_p \gamma^\mu d_r)$\\
\hline \hline
\etb
\captionsetup{singlelinecheck=off}
\caption[ . ]{Dimension-six operators other than the four-fermion ones~\cite{Grzadkowski:2010es}.
Following Ref.~\cite{Grzadkowski:2010es},  we adopt the notation 
$\vpj \equiv i \vp^\dag \left( D_\mu - \Db_\mu \right) \vp$, 
$\vpjt \equiv i \vp^\dag \left( \tau^I D_\mu - \Db_\mu \tau^I \right) \vp$, 
and  $\tvp = \eps \vp^*$, where $\eps = i \sigma_2$. 
Generation indices  $p,r$ are explicitly displayed for all fermion fields. 
The operator names  in the left column of each block should be supplemented with
generation indices of the fermion fields whenever necessary, e.g.,
$Q_{eW} \to Q_{eW}^{pr}$. Dirac, color, and weak isospin indices are always contracted
within the brackets, and not displayed. 
}
\label{tab:dim6-1}
\end{table}

\subsection{\it  Dimension six
\label{sect:dim6}}

Dimension six operators violating baryon number were first discussed in~\cite{Weinberg:1979sa,Wilczek:1979hc}. 
A first systematic classification  of all dimension-six operators was given in Ref.~\cite{Buchmuller:1985jz}. 
This issue was recently revisited by the authors of  Ref.~\cite{Grzadkowski:2010es}, who pointed out some 
redundancies in the classification of~\cite{Buchmuller:1985jz}. 
All in all, barring flavor structures and Hermitian conjugation,  the effective Lagrangian contains 
fifty-nine independent  dimension six B-conserving  operators:  fifteen do not contain any fermion fields, 
nineteen contain two fermion fields, and twenty-five contain four fermion fields.~\cite{Grzadkowski:2010es}. 
Similarly, there are five independent dimension-six operators that violate baryon number. 
All the dimension-six operators are  reported in Tables~\ref{tab:dim6-1} and \ref{tab:dim6-2}, 
taken from  Ref.~\cite{Grzadkowski:2010es}. 
Note that generation indices  $p,r,s,t$ are explicitly displayed for all fermion fields. 
The bosonic operators are all Hermitian.
For the operators containing fermions, Hermitian
conjugation is equivalent to transposition of generation indices in each of
the fermionic currents in classes $(\bar LL)(\bar LL)$, $(\bar RR)(\bar RR)$,
$(\bar LL)(\bar RR)$, and $\psi^2 \vp^2 D^2$ (except for $Q_{\vp u d}$).  For
the remaining operators with fermions, Hermitian conjugates are not listed
explicitly.

The B-conserving operators can be organized into three broad categories: 
\begin{itemize}
\item Operators that do not involve fermions: they  mainly modify couplings and properties of the gauge bosons 
(classes $X^3$, $\vp^4 D^2$,  $X^2\vp^2$)
and may affect reactions involving the physical Higgs boson. 

\item Operators that involve two fermion fields:  after EWSB  these induce 
corrections to the fermion mass matrices (class  $\psi^2\vp^3$), 
dipole moments  (both CP conserving and violating) of the fermions  (class  $\psi^2 X\vp$),   and corrections 
to the gauge-fermion couplings (class $\psi^2\vp^2 D$, through gauge bosons in the covariant derivatives). 

\item Four-fermion operators: these induce corrections to purely hadronic processes (four quarks), 
purely leptonic processes (four leptons), and semi-leptonic processes  (both charged- and neutral-current). 

\end{itemize} 

It is useful to recall here the  properties of the various operators under CP transformation. 
The bosonic operators containing $\wt X_{\mu\nu}$ are CP-odd, while the remaining ones are CP-even.
For a given  fermionic operator $Q$, the combinations $Q\ \mp \ Q^\dag$ is  CP-odd (-even). 
Therefore CP violation requires  a non-vanishing imaginary part
of the corresponding Wilson coefficient, when the operator is  expressed in the 
mass-eigenstate basis. 

Finally, note that both two- and four-fermion operators can mediate 
flavor-changing neutral current (FCNC) processes in both the lepton and quark sector. 
Within the SM  quark sector FCNCs are suppressed (arise only at loop level), while  leptonic 
FCNC are forbidden (lepton flavor  is a good quantum number). Even after minimally 
extending the SM to include neutrino masses,  leptonic FCNC amplitudes  are suppressed by 
the ratio $\Delta m_\nu^2 / M_W^2$, making lepton flavor violating (LFV) decays a very 
promising probe of BSM dynamics.

\begin{figure}[tb]
\centering 
\includegraphics[width=0.9\textwidth]{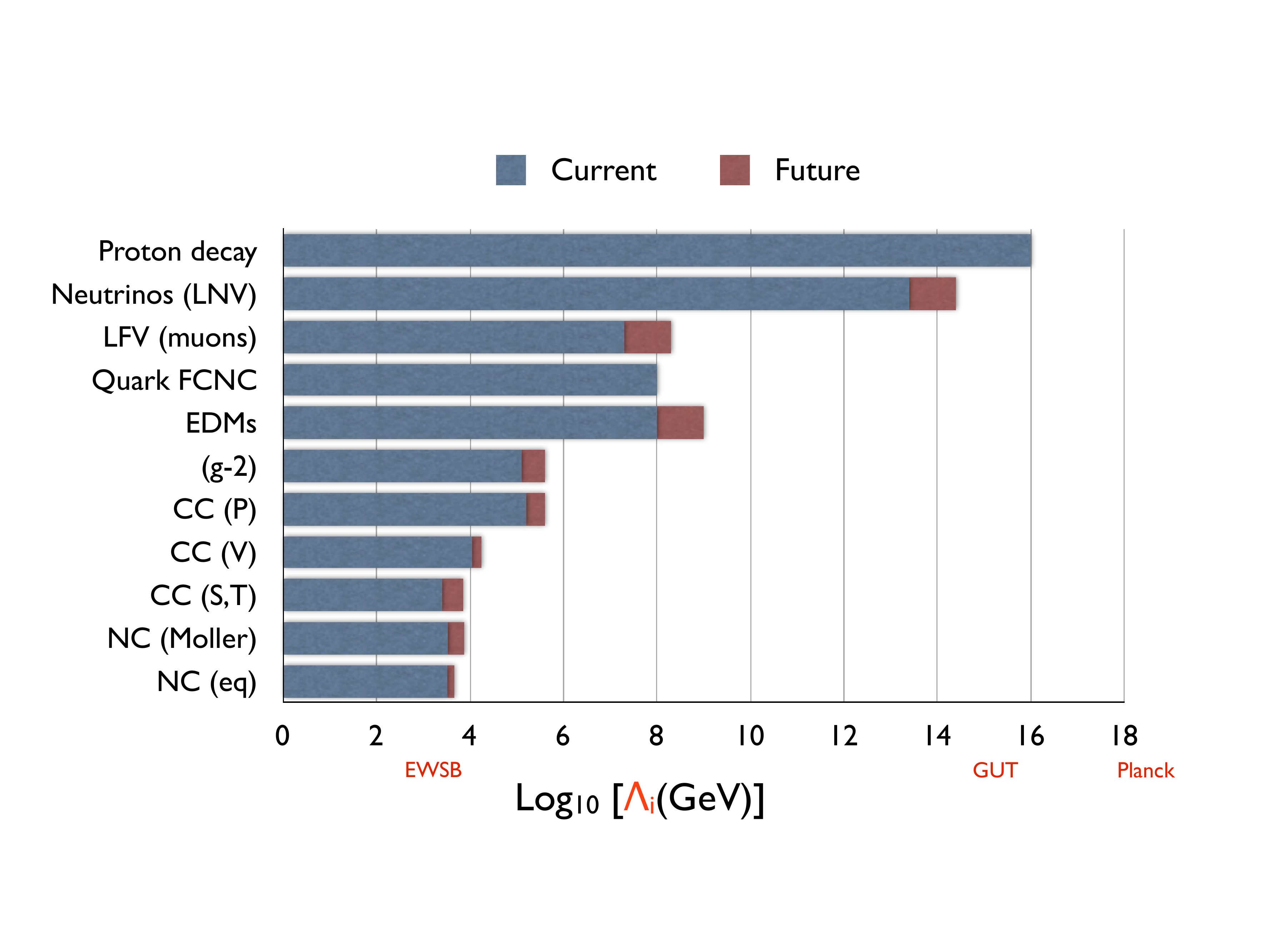}
\begin{minipage}[t]{16.5 cm}
\vspace{-1.cm}
\caption{Summary of current and future constraints on the 
effective new physics  scale $\Lambda_i$  defined in Eq.~(\ref{eq:Lambdadef}) 
arising from various low-energy observables. Note that the $\Lambda_i$ absorb
the Wilson coefficients and do not necessarily represent the masses of new degrees of freedom that become active at high energy scales.
 \label{fig:scales}}
\end{minipage}
\end{figure}

It is sometimes useful to define effective scales $\Lambda_i$ that absorb the Wilson coefficients. 
In the case of dimension-six operators one has: 
\be
\label{eq:Lambdadef}
\frac{1}{\Lambda_i^2} \approx  \frac{\alpha_i^{(6)}}{ \Lambda^2}\ \ \ ,
\ee
and similar definitions apply to the scale of operators of any dimensionality.
Bounds on the effective scale  $\Lambda_i$ 
of the dimension-5 and  dimension-6 operators can be obtained 
from a variety of low-energy and collider tests. These will be reviewed in detail in 
each of the chapters of this issue.  
Here we provide a first orientation,  summarizing the current and prospective bounds in Figure~\ref{fig:scales}. 

The effective scales $\Lambda_i$ that are indicated Fig.~\ref{fig:scales}  may be considered as the maximal scale probed by a given observable.
For example,  the bounds $\Lambda_{FCNC,CP} > 10^4$~TeV 
could be reconciled with TeV scale new particles, provided the new dynamics 
has approximate flavor (or CP) symmetries.  
Nevertheless, the effective scale $\Lambda_i$ is a good measure of how 
deeply a given measurement is constraining the new physics:  a large $\Lambda_i$ 
is either pushing the mass scale to large values or is telling us something deep about the symmetry structure 
of the TeV-scale BSM dynamics. 
With these caveats in mind,  Fig.~\ref{fig:scales} illustrates that 
the rare / forbidden processes provide the strongest bounds on the effective scale, 
with  $\Lambda_{B}  \sim 10^{16}$~GeV and  $\Lambda_{L}  \sim 10^{14}$~GeV. 
CP violation (EDMs and flavor sector) and  FCNC in both the quark and lepton sector 
provide the next  strongest  bounds on the effective scale,  namely $\Lambda_{CP}, \Lambda_{FCNC}  \sim  10^{4-5}$~TeV. 
Precision measurements such as the muon $g-2$ and $\pi \to e \nu$ provide constraints at the level of $\Lambda \sim 100$~TeV, 
while other charged-current and neutral-current probes are at the level of $\Lambda \sim 10$~TeV.
The reach of all these probes  overlaps with the LHC reach, so they will provide useful information to reconstruct possible new  
TeV-scale dynamics that might emerge at the LHC.

%
%The next strongest constraints arise from 
%precision EW observables form LEP, as well as CKM unitarity tests in beta decays,  
%with $\Lambda > O(10)$~TeV. 
%Next,  low-energy NC processes and other exotic CC structures imply bounds at the level of $\Lambda > 5$~TeV.
%

\begin{table}[t]
\centering
\renewcommand{\arraystretch}{1.5}
\begin{tabular}{||c|c||c|c||c|c||}
\hline\hline
\multicolumn{2}{||c||}{$(\bar LL)(\bar LL)$} & 
\multicolumn{2}{|c||}{$(\bar RR)(\bar RR)$} &
\multicolumn{2}{|c||}{$(\bar LL)(\bar RR)$}\\
\hline
$Q_{ll}$        & $(\bar l_p \gamma_\mu l_r)(\bar l_s \gamma^\mu l_t)$ &
$Q_{ee}$               & $(\bar e_p \gamma_\mu e_r)(\bar e_s \gamma^\mu e_t)$ &
$Q_{le}$               & $(\bar l_p \gamma_\mu l_r)(\bar e_s \gamma^\mu e_t)$ \\
$Q_{qq}^{(1)}$  & $(\bar q_p \gamma_\mu q_r)(\bar q_s \gamma^\mu q_t)$ &
$Q_{uu}$        & $(\bar u_p \gamma_\mu u_r)(\bar u_s \gamma^\mu u_t)$ &
$Q_{lu}$               & $(\bar l_p \gamma_\mu l_r)(\bar u_s \gamma^\mu u_t)$ \\
$Q_{qq}^{(3)}$  & $(\bar q_p \gamma_\mu \tau^I q_r)(\bar q_s \gamma^\mu \tau^I q_t)$ &
$Q_{dd}$        & $(\bar d_p \gamma_\mu d_r)(\bar d_s \gamma^\mu d_t)$ &
$Q_{ld}$               & $(\bar l_p \gamma_\mu l_r)(\bar d_s \gamma^\mu d_t)$ \\
$Q_{lq}^{(1)}$                & $(\bar l_p \gamma_\mu l_r)(\bar q_s \gamma^\mu q_t)$ &
$Q_{eu}$                      & $(\bar e_p \gamma_\mu e_r)(\bar u_s \gamma^\mu u_t)$ &
$Q_{qe}$               & $(\bar q_p \gamma_\mu q_r)(\bar e_s \gamma^\mu e_t)$ \\
$Q_{lq}^{(3)}$                & $(\bar l_p \gamma_\mu \tau^I l_r)(\bar q_s \gamma^\mu \tau^I q_t)$ &
$Q_{ed}$                      & $(\bar e_p \gamma_\mu e_r)(\bar d_s\gamma^\mu d_t)$ &
$Q_{qu}^{(1)}$         & $(\bar q_p \gamma_\mu q_r)(\bar u_s \gamma^\mu u_t)$ \\ 
&& 
$Q_{ud}^{(1)}$                & $(\bar u_p \gamma_\mu u_r)(\bar d_s \gamma^\mu d_t)$ &
$Q_{qu}^{(8)}$         & $(\bar q_p \gamma_\mu T^A q_r)(\bar u_s \gamma^\mu T^A u_t)$ \\ 
&& 
$Q_{ud}^{(8)}$                & $(\bar u_p \gamma_\mu T^A u_r)(\bar d_s \gamma^\mu T^A d_t)$ &
$Q_{qd}^{(1)}$ & $(\bar q_p \gamma_\mu q_r)(\bar d_s \gamma^\mu d_t)$ \\
&&&&
$Q_{qd}^{(8)}$ & $(\bar q_p \gamma_\mu T^A q_r)(\bar d_s \gamma^\mu T^A d_t)$\\
\hline\hline
\multicolumn{2}{||c||}{$(\bar LR)(\bar RL)$ and $(\bar LR)(\bar LR)$} &
\multicolumn{4}{|c||}{$B$-violating}\\\hline
$Q_{ledq}$ & $(\bar l_p^j e_r)(\bar d_s q_t^j)$ &
$Q_{duq}$ & \multicolumn{3}{|c||}{$\eps^{\alpha\beta\gamma} \eps_{jk} 
 \left[ (d^\alpha_p)^T C u^\beta_r \right]\left[(q^{\gamma j}_s)^T C l^k_t\right]$}\\
$Q_{quqd}^{(1)}$ & $(\bar q_p^j u_r) \eps_{jk} (\bar q_s^k d_t)$ &
$Q_{qqu}$ & \multicolumn{3}{|c||}{$\eps^{\alpha\beta\gamma} \eps_{jk} 
  \left[ (q^{\alpha j}_p)^T C q^{\beta k}_r \right]\left[(u^\gamma_s)^T C e_t\right]$}\\
$Q_{quqd}^{(8)}$ & $(\bar q_p^j T^A u_r) \eps_{jk} (\bar q_s^k T^A d_t)$ &
$Q_{qqq}^{(1)}$ & \multicolumn{3}{|c||}{$\eps^{\alpha\beta\gamma} \eps_{jk} \eps_{mn} 
  \left[ (q^{\alpha j}_p)^T C q^{\beta k}_r \right]\left[(q^{\gamma m}_s)^T C l^n_t\right]$}\\
$Q_{lequ}^{(1)}$ & $(\bar l_p^j e_r) \eps_{jk} (\bar q_s^k u_t)$ &
$Q_{qqq}^{(3)}$ & \multicolumn{3}{|c||}{$\eps^{\alpha\beta\gamma} (\tau^I \eps)_{jk} (\tau^I \eps)_{mn} 
  \left[ (q^{\alpha j}_p)^T C q^{\beta k}_r \right]\left[(q^{\gamma m}_s)^T C l^n_t\right]$}\\
$Q_{lequ}^{(3)}$ & $(\bar l_p^j \sigma_{\mu\nu} e_r) \eps_{jk} (\bar q_s^k \sigma^{\mu\nu} u_t)$ &
$Q_{duu}$ & \multicolumn{3}{|c||}{$\eps^{\alpha\beta\gamma} 
  \left[ (d^\alpha_p)^T C u^\beta_r \right]\left[(u^\gamma_s)^T C e_t\right]$}\\
\hline\hline
\end{tabular}
\caption{Dimension-six four-fermion operators from Ref.~\cite{Grzadkowski:2010es}.
Generation indices  $p,r,s,t$ are explicitly displayed for all fermion fields. 
The operator  names in the left column of each block should be supplemented with
generation indices of the fermion fields whenever necessary, e.g.,
$Q_{lq}^{(1)} \to Q_{lq}^{(1)prst}$. Dirac indices are always contracted
within the brackets, and not displayed. The same is true for the isospin and
colour indices in the upper part of the table.  In the lower-left
block of that table, colour indices are still contracted within the brackets,
while the isospin ones are made explicit. Colour indices are displayed only
for operators that violate the baryon number $B$ (lower-right block). 
\label{tab:dim6-2}}
\end{table}

\subsection{\it  Higher dimensional operators 
\label{sect:hdim}}

While most observables receive the leading BSM contribution 
from  dimension-six operators,  in some case it  is necessary to go beyond  
dimension six.    
For example, the leading contributions to Majorana neutrino  
transition magnetic moments arises from  dimension-seven operators~\cite{Bell:2006wi}. 
A different example requiring the need for higher dimensional operators 
involves the study of  non-standard flavor-changing  neutrino-matter interactions. 
The dimension-six contributions to these processes are highly suppressed because they 
are related by SU(2) gauge invariance to the corresponding charged LFV processes. 
This connection is lost at dimension eight, because  it is possible to construct  operators 
that contribute to neutrino-matter FCNC but not to charged LFV~\cite{Berezhiani:2001rs}.

To our knowledge, no complete classification of operators of dimension higher than six exists. 
However, a notable exception concerns  operators that violate total lepton number, 
which start at dimension five as discussed above. 
The  lepton number violating (LNV) operators up to and including dimension eleven have been classified  
in Ref.~\cite{deGouvea:2007xp}, that also studied the  contribution  of these 
operators to  various LNV processes, from low-energy to collider physics. 
Among the LNV operator, of particular interests are the $\Delta L=2$  six-fermion operators 
involving four quark field and two lepton fields that contribute to neutrino-less double beta decay 
($d d \to u u  e e$ at the quark-lepton level), that have also been 
studied in Ref.~\cite{Prezeau:2003xn}, although in a non SU(2) invariant framework. 
If the scale of LNV is close to the TeV scale, these operators can contribute to 
neutrino-less double beta decay rate at a level that will be probed in the 
next-generation experiments. 
The associated nuclear matrix elements are quite different from those arising in  the 
standard Majorana neutrino exchange and have been studied in Ref.~\cite{Prezeau:2003xn}.

%%%%%%%%%%%%%%%%%%%%%%%%%%%%%%%%%%%%%%%%%%%%%%%%%%%%%%%%
%        FIGURE
%%%%%%%%%%%%%%%%%%%%%%%%%%%%%%%%%%%%%%%%%%%%%%%%%%%%%%%%

\section{Discovery, Diagnosis, and Interpretability}
\label{sec:discover}

As discussed in Section \ref{sec:intro}, the processes of interest to this volume naturally fall into three broad categories: (a) rare and forbidden processes; (b) precision tests; and (c) cosmological and astrophysical probes. In the preceding sections of this introductory article, we have attempted to set a framework for the interpretation of results. The baseline is sufficiently reliable knowledge of SM expectations, often at the level of electroweak radiative corrections. The effects of BSM physics can be classified in a model-independent way in terms of effective operators of successively higher mass dimension with coefficients proportional to appropriate inverse powers of $\Lbsm$. Of course, for BSM dynamics involving new light degrees of freedom as in possible super-light gauge bosons that might account for the $a_\mu$ discrepancy, the effective operator analysis is not appropriate. Moreover, when the new degrees of freedom are not much heavier than the weak scale and enter at the loop level, the expansion of Eq.~(\ref{eq:EFT}) may not be optimal. The discussion of the supersymmetric loop contributions to charged current and neutral current observables in the corresponding reviews would constitute an example of the latter situation.

Broadly speaking, the three classes of low-energy studies provide new opportunities for significant discoveries, tools for diagnosing the detailed nature of BSM physics if it is observed, and/or placing strong constraints on various possibilities. The task of diagnosing and constraining BSM physics requires that one have in hand sufficient theoretical control over SM processes so that poorly-known aspects of the latter do not generate confusion about the former. We have already alluded to the QCD-related uncertainties that enter SM electroweak radiative corrections for CC and NC processes. The analogous strong interaction uncertainties that impact the interpretation of the muon anomalous magnetic moment are well-known. Here, we summarize the corresponding theoretical challenges as they bear on the prospects for discovery and diagnosis in the topics discussed in the remainder of this volume.

\begin{itemize}

\item[] {\bf Electric Dipole Moment Searches}. The observation of a non-zero permanent electric dipole moment (EDM) of a polar molecule, neutral atom, nucleon, or nucleus would represent a major discovery. The sensitivity of the next generation of experiments would not reach the level of sensitivity needed to probe the effects of SM CKM CP-violation. Consequently, a non-vanishing EDM would point to the effects of the QCD $\theta$-term and/or BSM CP-violation. As discussed above, the mass scale of BSM interactions probed in the next generation of EDM searches is approaching the tens of TeV level. 

The theoretical challenge for the EDM program is not to make room for discovery. It is, rather, to provide a path to diagnosing the underlying mechanism should a non-vanishing EDM be observed, or to set the framework for constraining possible sources of BSM CP-violation in the event of non-observation of EDMs in the next generation of experiments. At the level of the effective operators introduced here, one encounters twelve possible sources of BSM CP-violation involving the first generation quarks and leptons, photons, and gluons. These possible sources are summarized in Table I of the EDM review~\cite{Engel:2013lsa}. Polar molecules and paramagnetic atoms are in general sensitive to four different operators, though in practice are dominated by only two of the four: the electron EDM and a combination of the semileptonic four fermion operators: $Q_{\ell e d q}$ and $Q_{\ell e q u}^{(1)}$. Diamagnetic atoms, nucleons, and nuclei probe the $\theta$-term, quark EDMs, and six other dimension-six CPV operators. One component of the theoretical challenge is to identify a sufficient number of systems with complementary dependences on these operators to allow one to disentangle them should an EDM be observed in any one system. As discussed in the EDM review, a promising new direction in this regard are light nuclei whose EDMs may be probed in storage ring experiments.

The other theoretical challenge involves matching the underlying CPV operators onto hadronic, nuclear, and atomic degrees of freedom. While the theoretical uncertainties associated with many-body atomic computations appears to be reasonably small, the same cannot be said regarding the hadronic and nuclear uncertainties. In diamagnetic atoms such as $^{199}$Hg, the atomic EDM is most sensitive to the nuclear Schiff moment. The latter, in turn, arises from long-range time-reversal (TV) and parity-violating (PV) pion-exchange interactions that are induced by the non-leptonic CPV operators. Carrying out robust nuclear many-body computations of the Schiff moment remains a key theoretical challenge. The matching of the dimension six operators onto TVPV $\pi NN$ interactions and the nucleon EDMs constitutes a similar direction need of concerted theoretical effort. In some cases, the uncertainty associated with this matching  is an order of magnitude or larger, as discussed in the EDM review~\cite{Engel:2013lsa}. We direct the reader to Table 6 of that article, where a summary of \lq\lq best values" and  \lq\lq reasonable ranges" for the relevant hadronic matrix elements may be found. 

\item[] {\bf Probes of Lepton Number and Flavor Violation.} 
The physics responsible for neutrino masses and lepton mixing remains unknown, 
and it can be uniquely  probed  by searches for lepton flavor violation (LFV)  in charged-lepton processes  
($\mu \to e \gamma$,  $\mu \to e \bar{e} e$,  $\mu \to e$ conversion in nuclei, $\tau \to (e,\mu) \gamma$, $\tau \to (e, \mu) +$ hadrons, etc.), 
and lepton number violation (LNV)  in neutrino-less double beta decays.  
Given the tiny (for charged LFV) or null (for LNV)) SM background, these are ``discovery" searches, 
albeit experimentally extremely challenging.   Both  positive or null results from these searches will provide 
unique information about the symmetry structure of physics beyond the SM, whether or not the LHC detects new degrees of freedom  at the TeV scale. 

Charged LFV processes involving muons and tau are theoretically very clean: the atomic,  nuclear, 
(for $\mu \to e$ conversion),  and hadronic (for $\tau \to (e, \mu) +$ hadrons) effects are under reasonable control.
Therefore, charged LFV processes are not only discovery channels, but also powerful diagnostic tools. 
As an example, the physics reach of  $\mu \to e \gamma$,  $\mu \to e \bar{e} e$,  and $\mu \to e$ conversion in nuclei, 
and their complementarity are illustrated  in Figures 2 and 3 of the review  on lepton number and flavor violation~\cite{LFV-LNV}.
If observed, the relative rates of various $\mu \to e$ or $\tau \to (e,\mu)$  processes will provide 
information about the underlying mechanism (e.g. whether the dipole operator or other structures dominate).  
Similarly,  comparing   $\mu \to e$ with  $\tau \to (e,\mu)$ transitions will point to the structure of the underlying 
sources of flavor breaking. 

LNV is most sensitively probed by neutrino-less double beta decay.  
Observation of such process would indicate  that lepton number is not conserved and that neutrinos are their own antiparticles. 
Here the (inter-related) theoretical challenges are: (i)  diagnosing the underlying mechanism (light Majorana neutrino mediator, 
or new low-scale source of LNV);   (ii)  within  a given mechanism, reducing the uncertainty in the  nuclear matrix elements, 
which is key to extracting bounds on the underlying model parameters (for example  $\langle m_{\beta \beta} \rangle$ 
if the mechanism is the exchange of a light Majorana neutrino). 
Both these issues are discussed at depth in Ref.~\cite{LFV-LNV}, and the status of nuclear matrix elements is summarized in their Figure 8.

\item[] {\bf Charged and Neutral Current Processes}. The emphasis of precise studies of weak decays and neutral current processes lies on diagnosing key aspects of possible BSM physics, should a discovery be made at the energy frontier. Conversely, agreement of these studies with SM expectations can place severe constraints on BSM scenarios. Within the Minimal Supersymmetric Standard Model, for example, a comparison of weak charges of the proton and electron as measured in PV electron scattering could yield information as to whether the MSSM violates or respects $B-L$ (see Figure 10 of the NC review~\cite{neutral-current}). Alternately, a comparison of the results of first row CKM unitarity tests with those of pion leptonic decays could yield information on the spectrum of supersymmetric particles (see Figure 6 of the weak decay review). Studies of the PV asymmetries could also yield insights into the nature of new neutral gauge bosons ($Z^\prime$), such as in string-theory motivated scenarios. In each case, the low-energy studies have the potential to complement what one may learn from collider studies. Deep inelastic PV electron-deuteron scattering has a unique sensitivity to a light,  \lq\lq leptophobic" $Z^\prime$ that is a candidate for explaining recent results from the Tevatron. Conversely, if new particles are too heavy to be produced directly at the LHC or other colliders, the effective operator description would apply equally to the interpretation of collider results and low-energy tests. In this case, one can see directly the power of low-energy measurements to complement the reach at the energy frontier, as illustrated in Figures 3 and 4 of the weak decay review~\cite{charged-current}.

Discerning the effects of these BSM scenarios will require not only pushing the level of experimental sensitivity to the next decade, but also making similar improvements in the reliability of the theoretical SM inputs. Among the primary challenges are the $W\gamma$ and $Z\gamma$ box graphs that enter the extraction of $V_{ud}$ from neutron and nuclear $\beta$-decay and the proton weak charge from elastic PV electron-proton scattering. The interpretation of neutron and nuclear $\beta$-decay correlations in terms of non-$(V-A)\otimes (V-A)$ interactions similarly calls for refined computations of the nucleon scalar and tensor form factors (see Figure 3 of the weak decay review~\cite{charged-current}).  Determining the magnitude of higher-twist corrections and contributions associated with charge symmetry-violation in parton distribution functions would enhance the BSM diagnostic power of PV deep inelastic electron scattering.

\item[] {\bf Neutrinos: Terrestrial,  Astrophysical, and Cosmological Probes.}
Neutrinos probe a rich sector of the BSM dynamics, largely inaccessible at the high-energy frontier. 
In the EFT  language used in  this review, neutrino experiments  probe both the  particle content 
and symmetry structure of the EFT, through question such as:   are there  light sterile  right-handed neutrinos?  How many? 
If so, do they acquire a Majorana mass term (dimiension-three lepton-number violating operator), 
and do they have Yukawa interactions with Higgs and left-handed leptons (dimension-four operator)? 
Is there a direct Majorana mass term for left-handed neutrinos  (dimension-five operator of  Eq.~(\ref{eq:mnumajorana}))?
Is CP symmetry violated by the above couplings? 

Regardless the origin of the light neutrino  mass matrix (Dirac or Majorana), as discussed in Ref.~\cite{Balantekin:2013tqa},
through oscillation experiments involving solar, atmospheric, reactor, and accelerator neutrinos  
we  know  a great deal about the mixing angles and mass-splitting of 
the three active neutrinos.  In terms of determining neutrino properties, besides the 
Majorana or Dirac nature,  the remaining  
open questions concern the overall mass scale, the mass hierarchy, 
and CP violation in the mixing matrix, and  the possibility of sterile states or non-standard neutrino-matter interactions. 
It is also worth pointing out that now it has become possible to use neutrinos as quantitative astrophysical probes of the Sun 
and the Standard Solar Model,  as discussed in the neutrino oscillation review~\cite{Balantekin:2013tqa}.

Last but not least,  as discussed in the review on neutrinos in cosmology and astrophysics~\cite{baha-george},  
neutrinos shape key aspects of the  Early Universe and core collapse supernovae dynamics, 
because they carry a dominant fraction of the total energy and entropy in these environments.  
Given the energy and flavor-dependence of neutrino processes in the EU or SN,  
the key open theoretical challenge involves 
understanding neutrino flavor transformation in medium (including coherent and incoherent scattering). 
A robust understanding of this problem will allow us to
(i)  combine  recent and future developments in observational cosmology and neutrino experiment to 
probe the presence of new physics in the neutrino sector (e.g. sterile neutrinos); 
(ii) reliably predict the flavor and spectral composition of SN neutrino signal; 
(iii) explore  the origin of the lightest and heaviest nuclei, through nucleosynthesis in the early universe 
and astrophysical sites.

\item[]{\bf Dark Matter and Nuclear Physics.} Dark matter (DM) provides approximately one-fifth of the mass-energy content of the Universe  and the SM has no candidate for it.  Several theoretically well-motivated SM extensions contain DM candidates. 
However, the constraints of relic density,  stability (on cosmological time scales), and extremely weak coupling 
to ordinary matter still leave many open possibilities.  
While ``WIMPs" (weakly interacting massive particles)
remain a very attractive candidate for DM,  as illustrated in Figure 1 of the review article~\cite{dark-matter},  
the mass and interaction strengths of  DM candidates span fifty orders of magnitude!  Reference~\cite{dark-matter} discusses how nuclear physics can play an important role in both  WIMP  and non-WIMP DM searches.  For WIMP  direct detection searches, one obvious challenge for nuclear theory is the  calculation of WIMP-nucleus cross sections starting from quark-WIMP interactions. 
Similarly,   facilities dedicated to nuclear physics are well-poised to investigate certain non-WIMP models,  such as ``Hidden Sector Models".    In parallel to this, developments in observational cosmology permit probes of the relativistic energy density at early epochs and thus provide new ways to constrain dark-matter models, provided nuclear physics inputs are sufficiently well-known. The emerging confluence of accelerator, astrophysical, and cosmological constraints permit searches for dark-matter candidates in a greater range of masses and interaction strengths than heretofore possible.

\item[]{\bf Hadronic Parity-Violation}. Apart from these BSM probes, low-energy electroweak interactions continue to provide unique windows on poorly understood aspects of nucleon and nuclear structure. In this series of reviews we have not emphasized this aspect of the low-energy program, but have included one article in this spirit focusing on hadronic PV~\cite{hadronicPV}. The nonleptonic weak interaction in the SM remains a topic of considerable interest as it challenges our understanding of the interplay of strong and weak interactions. More broadly, several puzzles remain to be addressed, such as the origin of the $\Delta I=1/2$ rule in non-leptonic weak decays of mesons; the tension between $S$- and $P$-wave amplitudes in strangeness changing, non-leptonic hyperon decays; and the enhanced PV asymmetries in strangeness-changing radiative hyperon decays. Whether the origin of these puzzles lies in the role played by the strange quark in hadronic dynamics of something not yet identified at the interface of quark-quark weak interactions and non-perturbative QCD remains an open question. 

The study of the strangeness conserving hadronic weak interaction (HWI) provides a probe of these underlying dynamics without the presence of the strange quark. Theoretically, one  may characterize the HWI in terms of a low-energy effective field theory involving nucleon and -- for appropriate energy scales -- pion degrees of 
freedom. 
%\cite{Zhu:2004vw,RamseyMusolf:2006dz,Girlanda:2008ts,Phillips:2008hn}. 
An alternate, model-dependent approach based on meson-exchange interactions has been widely used to interpret the experimental  results
% \cite{Desplanques:1979hn,Adelberger:1985ik} 
(a detailed translation between the two approaches appears in Ref.~\cite{hadronicPV}). Either way, the goal of the combined experimental and theoretical program is to obtain from measurements values for the hadronic-level parameters that may then be matched onto the underlying non-leptonic, strangeness conserving HWI. The review of hadronic PV outlines progress and opportunities in this program~\cite{hadronicPV}. In particular, results of an updated global analysis in terms of the meson-exchange model parameters appears in Figure 3 of Ref.~\cite{hadronicPV}. 
The matching of the effective hadronic parameters onto the quark-level HWI remains an open theoretical challenge.

\end{itemize}

\section{Conclusion}

We now leave it to the reader to delve into the details of each of the aforementioned directions. Suffice it to say, the program of low-energy probes of  BSM physics and related topics  is a rich,  diverse, and multidisciplinary field of research. These studies provide a powerful and unique window on dynamics at high energy scales and have the potential to uncover key aspects of new laws of nature that may have been more apparent in the early universe than they are today. Their physics reach complements that of experiments at the energy and cosmic frontiers. They present a number of theoretically compelling challenges. We hope that this series of articles will give the reader a taste of the opportunities and excitement that the field engenders.

%%%%%%%%%%%%%%%%%%%%%%%%%%%%%%%%%%%%%%%%%%%%%%%%%%%%%%%%
%       BIBLIOGRAPHY  (WITH  BIBTEX) 
%%%%%%%%%%%%%%%%%%%%%%%%%%%%%%%%%%%%%%%%%%%%%%%%%%%%%%%%
\bibliographystyle{doiplain}
\bibliography{Vincenzo}

\begin{thebibliography}{10}

\bibitem{Erler:2004cx}
Jens Erler and Michael~J. Ramsey-Musolf.
\newblock {Low energy tests of the weak interaction}.
\newblock {\em Prog.Part.Nucl.Phys.}, 54:351\unskip--\ignorespaces 442, 2005,
  \doi{10.1016/j.ppnp.2004.08.001}, \eprint{arXiv}{hep-ph/0404291}.

\bibitem{RamseyMusolf:2006vr}
M.J. Ramsey-Musolf and S.~Su.
\newblock {Low Energy Precision Test of Supersymmetry}.
\newblock {\em Phys.Rept.}, 456:1\unskip--\ignorespaces 88, 2008,
  \doi{10.1016/j.physrep.2007.10.001}, \eprint{arXiv}{hep-ph/0612057}.

\bibitem{Pospelov:2005pr}
Maxim Pospelov and Adam Ritz.
\newblock {Electric dipole moments as probes of new physics}.
\newblock {\em Annals Phys.}, 318:119\unskip--\ignorespaces 169, 2005,
  \doi{10.1016/j.aop.2005.04.002}, \eprint{arXiv}{hep-ph/0504231}.

\bibitem{Elliott:2002xe}
Steven~R. Elliott and Petr Vogel.
\newblock {Double beta decay}.
\newblock {\em Ann.Rev.Nucl.Part.Sci.}, 52:115\unskip--\ignorespaces 151, 2002,
  \doi{10.1146/annurev.nucl.52.050102.090641}, \eprint{arXiv}{hep-ph/0202264}.

\bibitem{Elliott:2004hr}
Steven~R. Elliott and Jonathan Engel.
\newblock {Double beta decay}.
\newblock {\em J.Phys.}, G30:R183, 2004, \doi{10.1088/0954-3899/30/9/R01},
  \eprint{arXiv}{hep-ph/0405078}.

\bibitem{Avignone:2007fu}
III Avignone, Frank~T., Steven~R. Elliott, and Jonathan Engel.
\newblock {Double Beta Decay, Majorana Neutrinos, and Neutrino Mass}.
\newblock {\em Rev.Mod.Phys.}, 80:481\unskip--\ignorespaces 516, 2008,
  \doi{10.1103/RevModPhys.80.481}, \eprint{arXiv}{0708.1033}.

\bibitem{Herczeg:2001vk}
P.~Herczeg.
\newblock {Beta decay beyond the standard model}.
\newblock {\em Prog.Part.Nucl.Phys.}, 46:413\unskip--\ignorespaces 457, 2001,
  \doi{10.1016/S0146-6410(01)00149-1}.

\bibitem{Severijns:2006dr}
Nathal Severijns, Marcus Beck, and Oscar Naviliat-Cuncic.
\newblock {Tests of the standard electroweak model in beta decay}.
\newblock {\em Rev.Mod.Phys.}, 78:991\unskip--\ignorespaces 1040, 2006,
  \doi{10.1103/RevModPhys.78.991}, \eprint{arXiv}{nucl-ex/0605029}.

\bibitem{Severijns:2011zz}
Nathal Severijns and Oscar Naviliat-Cuncic.
\newblock {Symmetry tests in nuclear beta decay}.
\newblock {\em Ann.Rev.Nucl.Part.Sci.}, 61:23\unskip--\ignorespaces 46, 2011,
  \doi{10.1146/annurev-nucl-102010-130410}.

\bibitem{Kumar:2013yoa}
K.S. Kumar, Sonny Mantry, W.J. Marciano, and P.A. Souder.
\newblock {Low Energy Measurements of the Weak Mixing Angle}.
\newblock 2013, \eprint{arXiv}{1302.6263}.

\bibitem{Musolf:1993tb}
M.J. Musolf, T.W. Donnelly, J.~Dubach, S.J. Pollock, S.~Kowalski, et~al.
\newblock {Intermediate-energy semileptonic probes of the hadronic neutral
  current}.
\newblock {\em Phys.Rept.}, 239:1\unskip--\ignorespaces 178, 1994,
  \doi{10.1016/0370-1573(94)90040-X}.

\bibitem{Adelberger:1985ik}
E.G. Adelberger and W.C. Haxton.
\newblock {Parity Violation in the Nucleon-Nucleon Interaction}.
\newblock {\em Ann.Rev.Nucl.Part.Sci.}, 35:501\unskip--\ignorespaces 558, 1985.

\bibitem{RamseyMusolf:2006dz}
Michael~J. Ramsey-Musolf and Shelley~A. Page.
\newblock {Hadronic parity violation: A New view through the looking glass}.
\newblock {\em Ann.Rev.Nucl.Part.Sci.}, 56:1\unskip--\ignorespaces 52, 2006,
  \doi{10.1146/annurev.nucl.54.070103.181255}, \eprint{arXiv}{hep-ph/0601127}.

\bibitem{Sirlin:1977sv}
A.~Sirlin.
\newblock {Current Algebra Formulation of Radiative Corrections in Gauge
  Theories and the Universality of the Weak Interactions}.
\newblock {\em Rev.Mod.Phys.}, 50:573, 1978, \doi{10.1103/RevModPhys.50.573}.

\bibitem{Marciano:1985pd}
W.J. Marciano and A.~Sirlin.
\newblock {Radiative Corrections to beta Decay and the Possibility of a Fourth
  Generation}.
\newblock {\em Phys.Rev.Lett.}, 56:22, 1986, \doi{10.1103/PhysRevLett.56.22}.

\bibitem{Marciano:1993sh}
William~J. Marciano and A.~Sirlin.
\newblock {Radiative corrections to pi(lepton 2) decays}.
\newblock {\em Phys.Rev.Lett.}, 71:3629\unskip--\ignorespaces 3632, 1993,
  \doi{10.1103/PhysRevLett.71.3629}.

\bibitem{Marciano:2005ec}
William~J. Marciano and Alberto Sirlin.
\newblock {Improved calculation of electroweak radiative corrections and the
  value of {$V_{ud}$}}.
\newblock {\em Phys.Rev.Lett.}, 96:032002, 2006,
  \doi{10.1103/PhysRevLett.96.032002}, \eprint{arXiv}{hep-ph/0510099}.

\bibitem{Ivanov:2012qe}
A.N. Ivanov, M.~Pitschmann, and N.I. Troitskaya.
\newblock {Neutron Beta-Decay as Laboratory for Test of Standard Model}.
\newblock 2012, \eprint{arXiv}{1212.0332}.

\bibitem{charged-current}
Vincenzo Cirigliano, Susan~V. Gardner, and Barry~R. Holstein.
\newblock {Beta Decays and Non-Standard Interactions in the LHC Era}.
\newblock 2013, \eprint{arXiv}{1303.6953}.

\bibitem{Marciano:1980pb}
W.J. Marciano and A.~Sirlin.
\newblock {Radiative Corrections to Neutrino Induced Neutral Current Phenomena
  in the SU(2)-L x U(1) Theory}.
\newblock {\em Phys.Rev.}, D22:2695, 1980, \doi{10.1103/PhysRevD.31.213,
  10.1103/PhysRevD.22.2695}.

\bibitem{Marciano:1982mm}
W.J. Marciano and A.~Sirlin.
\newblock {RADIATIVE CORRECTIONS TO ATOMIC PARITY VIOLATION}.
\newblock {\em Phys.Rev.}, D27:552, 1983, \doi{10.1103/PhysRevD.27.552}.

\bibitem{Sarantakos:1982bp}
S.~Sarantakos, A.~Sirlin, and W.J. Marciano.
\newblock {Radiative Corrections to Neutrino-Lepton Scattering in the SU(2)-L x
  U(1) Theory}.
\newblock {\em Nucl.Phys.}, B217:84, 1983, \doi{10.1016/0550-3213(83)90079-2}.

\bibitem{Marciano:1983ss}
W.J. Marciano and A.~Sirlin.
\newblock {On Some General Properties of the O(alpha) Corrections to Parity
  Violation in Atoms}.
\newblock {\em Phys.Rev.}, D29:75, 1984, \doi{10.1103/PhysRevD.29.75,
  10.1103/PhysRevD.31.213.2}.

\bibitem{veltman}
M.J.G. Veltman.
\newblock {Limit on Mass Differences in the Weinberg Model}.
\newblock {\em Nucl.Phys.}, B123:89, 1977, \doi{10.1016/0550-3213(77)90342-X}.

\bibitem{neutral-current}
Jens Erler and Shufang Su.
\newblock {The Weak Neutral Current}.
\newblock 2013, \eprint{arXiv}{1303.5522}.

\bibitem{Czarnecki:1995fw}
Andrzej Czarnecki and William~J. Marciano.
\newblock {Electroweak radiative corrections to polarized Moller scattering
  asymmetries}.
\newblock {\em Phys.Rev.}, D53:1066\unskip--\ignorespaces 1072, 1996,
  \doi{10.1103/PhysRevD.53.1066}, \eprint{arXiv}{hep-ph/9507420}.

\bibitem{Erler:2004in}
Jens Erler and Michael~J. Ramsey-Musolf.
\newblock {The Weak mixing angle at low energies}.
\newblock {\em Phys.Rev.}, D72:073003, 2005, \doi{10.1103/PhysRevD.72.073003},
  \eprint{arXiv}{hep-ph/0409169}.

\bibitem{zeldovich}
Ya.B. Zeldovich.
\newblock {\em Sov.Phys. JETP}, 6:1184, 1958.

\bibitem{Musolf:1990sa}
M.J. Musolf and Barry~R. Holstein.
\newblock {Observability of the anapole moment and neutrino charge radius}.
\newblock {\em Phys.Rev.}, D43:2956\unskip--\ignorespaces 2970, 1991,
  \doi{10.1103/PhysRevD.43.2956}.

\bibitem{Kurylov:2003zh}
A.~Kurylov, M.J. Ramsey-Musolf, and S.~Su.
\newblock {Probing supersymmetry with parity violating electron scattering}.
\newblock {\em Phys.Rev.}, D68:035008, 2003, \doi{10.1103/PhysRevD.68.035008},
  \eprint{arXiv}{hep-ph/0303026}.

\bibitem{Peskin:1990zt}
Michael~E. Peskin and Tatsu Takeuchi.
\newblock {A New constraint on a strongly interacting Higgs sector}.
\newblock {\em Phys.Rev.Lett.}, 65:964\unskip--\ignorespaces 967, 1990,
  \doi{10.1103/PhysRevLett.65.964}.

\bibitem{Golden:1990ig}
Mitchell Golden and Lisa Randall.
\newblock {RADIATIVE CORRECTIONS TO ELECTROWEAK PARAMETERS IN TECHNICOLOR
  THEORIES}.
\newblock {\em Nucl.Phys.}, B361:3\unskip--\ignorespaces 23, 1991,
  \doi{10.1016/0550-3213(91)90614-4}.

\bibitem{Marciano:1990dp}
William~J. Marciano and Jonathan~L. Rosner.
\newblock {Atomic parity violation as a probe of new physics}.
\newblock {\em Phys.Rev.Lett.}, 65:2963\unskip--\ignorespaces 2966, 1990,
  \doi{10.1103/PhysRevLett.65.2963}.

\bibitem{Kennedy:1990ib}
D.C. Kennedy and Paul Langacker.
\newblock {Precision electroweak experiments and heavy physics: A Global
  analysis}.
\newblock {\em Phys.Rev.Lett.}, 65:2967\unskip--\ignorespaces 2970, 1990,
  \doi{10.1103/PhysRevLett.65.2967}.

\bibitem{Kennedy:1991sn}
D.C. Kennedy and Paul Langacker.
\newblock {Precision electroweak experiments and heavy physics: An Update}.
\newblock {\em Phys.Rev.}, D44:1591\unskip--\ignorespaces 1592, 1991,
  \doi{10.1103/PhysRevD.44.1591}.

\bibitem{Altarelli:1990zd}
Guido Altarelli and Riccardo Barbieri.
\newblock {Vacuum polarization effects of new physics on electroweak
  processes}.
\newblock {\em Phys.Lett.}, B253:161\unskip--\ignorespaces 167, 1991,
  \doi{10.1016/0370-2693(91)91378-9}.

\bibitem{Holdom:1990tc}
B.~Holdom and J.~Terning.
\newblock {Large corrections to electroweak parameters in technicolor
  theories}.
\newblock {\em Phys.Lett.}, B247:88\unskip--\ignorespaces 92, 1990,
  \doi{10.1016/0370-2693(90)91054-F}.

\bibitem{Hagiwara:1994pw}
Kaoru Hagiwara, S.~Matsumoto, D.~Haidt, and C.S. Kim.
\newblock {A Novel approach to confront electroweak data and theory}.
\newblock {\em Z.Phys.}, C64:559\unskip--\ignorespaces 620, 1994,
  \doi{10.1007/BF01957770}, \eprint{arXiv}{hep-ph/9409380}.

\bibitem{Erler:2003yk}
Jens Erler, Andriy Kurylov, and Michael~J Ramsey-Musolf.
\newblock {The Weak charge of the proton and new physics}.
\newblock {\em Phys.Rev.}, D68:016006, 2003, \doi{10.1103/PhysRevD.68.016006},
  \eprint{arXiv}{hep-ph/0302149}.

\bibitem{Appelquist:1993ka}
Thomas Appelquist and Guo-Hong Wu.
\newblock {The Electroweak chiral Lagrangian and new precision measurements}.
\newblock {\em Phys.Rev.}, D48:3235\unskip--\ignorespaces 3241, 1993,
  \doi{10.1103/PhysRevD.48.3235}, \eprint{arXiv}{hep-ph/9304240}.

\bibitem{Longhitano:1980iz}
Anthony~C. Longhitano.
\newblock {Heavy Higgs Bosons in the Weinberg-Salam Model}.
\newblock {\em Phys.Rev.}, D22:1166, 1980, \doi{10.1103/PhysRevD.22.1166}.

\bibitem{Feruglio:1992wf}
F.~Feruglio.
\newblock {The Chiral approach to the electroweak interactions}.
\newblock {\em Int.J.Mod.Phys.}, A8:4937\unskip--\ignorespaces 4972, 1993,
  \doi{10.1142/S0217751X93001946}, \eprint{arXiv}{hep-ph/9301281}.

\bibitem{Wudka:1994ny}
Jose Wudka.
\newblock {Electroweak effective Lagrangians}.
\newblock {\em Int.J.Mod.Phys.}, A9:2301\unskip--\ignorespaces 2362, 1994,
  \doi{10.1142/S0217751X94000959}, \eprint{arXiv}{hep-ph/9406205}.

\bibitem{Weinberg:1979sa}
Steven Weinberg.
\newblock {Baryon and Lepton Nonconserving Processes}.
\newblock {\em Phys.Rev.Lett.}, 43:1566\unskip--\ignorespaces 1570, 1979,
  \doi{10.1103/PhysRevLett.43.1566}.

\bibitem{Grzadkowski:2010es}
B.~Grzadkowski, M.~Iskrzynski, M.~Misiak, and J.~Rosiek.
\newblock {Dimension-Six Terms in the Standard Model Lagrangian}.
\newblock {\em JHEP}, 1010:085, 2010, \doi{10.1007/JHEP10(2010)085},
  \eprint{arXiv}{1008.4884}.

\bibitem{Wilczek:1979hc}
Frank Wilczek and A.~Zee.
\newblock {Operator Analysis of Nucleon Decay}.
\newblock {\em Phys.Rev.Lett.}, 43:1571\unskip--\ignorespaces 1573, 1979,
  \doi{10.1103/PhysRevLett.43.1571}.

\bibitem{Buchmuller:1985jz}
W.~Buchmuller and D.~Wyler.
\newblock {Effective Lagrangian Analysis of New Interactions and Flavor
  Conservation}.
\newblock {\em Nucl. Phys.}, B268:621, 1986,
  \doi{10.1016/0550-3213(86)90262-2}.

\bibitem{Bell:2006wi}
Nicole~F. Bell, Mikhail Gorchtein, Michael~J. Ramsey-Musolf, Petr Vogel, and
  Peng Wang.
\newblock {Model independent bounds on magnetic moments of Majorana neutrinos}.
\newblock {\em Phys.Lett.}, B642:377\unskip--\ignorespaces 383, 2006,
  \doi{10.1016/j.physletb.2006.09.055}, \eprint{arXiv}{hep-ph/0606248}.

\bibitem{Berezhiani:2001rs}
Zurab Berezhiani and Anna Rossi.
\newblock {Limits on the nonstandard interactions of neutrinos from e+ e-
  colliders}.
\newblock {\em Phys.Lett.}, B535:207\unskip--\ignorespaces 218, 2002,
  \doi{10.1016/S0370-2693(02)01767-7}, \eprint{arXiv}{hep-ph/0111137}.

\bibitem{deGouvea:2007xp}
Andre de~Gouvea and James Jenkins.
\newblock {A Survey of Lepton Number Violation Via Effective Operators}.
\newblock {\em Phys.Rev.}, D77:013008, 2008, \doi{10.1103/PhysRevD.77.013008},
  \eprint{arXiv}{0708.1344}.

\bibitem{Prezeau:2003xn}
Gary Prezeau, M.~Ramsey-Musolf, and Petr Vogel.
\newblock {Neutrinoless double beta decay and effective field theory}.
\newblock {\em Phys.Rev.}, D68:034016, 2003, \doi{10.1103/PhysRevD.68.034016},
  \eprint{arXiv}{hep-ph/0303205}.

\bibitem{Engel:2013lsa}
Jonathan Engel, Michael~J. Ramsey-Musolf, and U.~van Kolck.
\newblock {Electric Dipole Moments of Nucleons, Nuclei, and Atoms: The Standard
  Model and Beyond}.
\newblock 2013, \eprint{arXiv}{1303.2371}.

\bibitem{LFV-LNV}
Andre de~Gouvea and Petr Vogel.
\newblock {Lepton Flavor and Number Conservation, and Physics Beyond the
  Standard Model}.
\newblock 2013, \eprint{arXiv}{1303.4097}.

\bibitem{Balantekin:2013tqa}
A.B. Balantekin and W.C. Haxton.
\newblock {Neutrino Oscillations}.
\newblock 2013, \eprint{arXiv}{1303.2272}.

\bibitem{baha-george}
A.B. Balantekin and G.M. Fuller.
\newblock {Neutrinos in Cosmology and Astrophysics}.
\newblock 2013, \eprint{arXiv}{1303.3874}.

\bibitem{dark-matter}
Susan Gardner and George Fuller.
\newblock {Dark Matter Studies Entrain Nuclear Physics}.
\newblock 2013, \eprint{arXiv}{1303.4758}.

\bibitem{hadronicPV}
W.C. Haxton and B.R. Holstein.
\newblock {Hadronic Parity Violation}.
\newblock 2013, \eprint{arXiv}{1303.4132}.

\end{thebibliography}

\end{document}